\documentclass[%
 reprint,
 superscriptaddress,
 nofootinbib,
 amsmath,amssymb,
 aps,
 pra
]{revtex4-1}

\usepackage[pdftex]{graphicx}
\usepackage[pdftex,
            colorlinks=true,
            citecolor=blue,
            linkcolor=blue]{hyperref}
\usepackage{bm}
\usepackage{braket}
\DeclareMathOperator{\sgn}{sgn}
\DeclareMathOperator{\Pf}{Pf}

\begin{document}

\title{Classification of topological crystalline superconducting nodes on high-symmetry lines: \texorpdfstring{\\}{}
Point nodes, line nodes, and Bogoliubov Fermi surfaces}

\author{Shuntaro Sumita}
\email[]{s.sumita@scphys.kyoto-u.ac.jp}
\affiliation{%
 Department of Physics, Kyoto University, Kyoto 606-8502, Japan
}%

\author{Takuya Nomoto}
\affiliation{%
 Department of Applied Physics, The University of Tokyo, Tokyo 113-8656, Japan
}%

\author{Ken Shiozaki}
\affiliation{%
 Yukawa Institute for Theoretical Physics, Kyoto University, Kyoto 606-8502, Japan
}%

\author{Youichi Yanase}
\affiliation{%
 Department of Physics, Kyoto University, Kyoto 606-8502, Japan
}%


\date{\today}

\begin{abstract}
Recent development in exact classification of a superconducting gap has elucidated various unconventional gap structures, which have not been predicted by the classification of order parameter based on the point group.
One of the important previous results is that all symmetry-protected line nodes are characterized by nontrivial topological numbers.
Another intriguing discovery is the gap structures depending on the angular momentum $j_z$ of normal Bloch states on threefold and sixfold rotational-symmetric lines in the Brillouin zone.
Stimulated by these findings, we classify irreducible representations of the Bogoliubov-de Gennes Hamiltonian at each $\bm{k}$ point on a high-symmetry $n$-fold ($n = 2$, $3$, $4$, and $6$) axis for centrosymmetric and paramagnetic superconductors, by using the combination of group theory and $K$ theory.
This leads to the classification of all crystal symmetry-protected nodes (including $j_z$-dependent nodes) on the axis that crosses a normal-state Fermi surface.
As a result, it is shown that the classification by group theory completely corresponds with the topological classification.
Based on the obtained results, we discuss superconducting gap structures in SrPtAs, CeCoIn$_5$, UPt$_3$, and UCoGe.
\end{abstract}

\maketitle


\section{Introduction}
Triggered by the discovery of unconventional superconductors, classification of superconductivity by symmetry has been started from 1980s~\cite{Volovik1984, Volovik1985, Anderson1984}.
Such theories give classification of superconducting \textit{order parameter} based on the crystal point group, which was summarized by Sigrist and Ueda~\cite{Sigrist-Ueda}.
Although their classification has been used for analyses of excitation spectrum, it may not provide a precise result of the superconducting \textit{gap}, namely an excitation energy in the Bogoliubov quasiparticle spectrum.
Indeed, several studies have shown that nonsymmorphic crystalline symmetry induces unconventional gap structures which are not predicted by the classification of order parameter~\cite{Yarzhemsky1992, Norman1995, Yarzhemsky1998, Yarzhemsky2000, Yarzhemsky2003, Yarzhemsky2008, Micklitz2009, Kobayashi2016, Yanase2016, Nomoto2016_PRL, Micklitz2017_PRB, Nomoto2017, Micklitz2017_PRL, Sumita2017, Sumita2018, Kobayashi2018}.
That is because the space group symmetry is not taken into account in the above classification method; nonsymmorphic symmetry is neglected by the classification based on the point group.
Furthermore, our recent group-theoretical researches have suggested unpredicted gap structures even in symmorphic systems; on threefold and sixfold rotational-symmetric lines in the Brillouin zone (BZ), the presence or absence of superconducting nodes depends on the total angular momentum of normal Bloch states $j_z$ (called $j_z$-dependent gap structures in this paper)~\cite{Nomoto2016_PRB, Sumita2018}.

Group-theoretical classification for the gap is closely related to the existence of a topological number.
Indeed, all types of line nodes on mirror- or glide-invariant planes in the BZ, which were completely elucidated by group theory~\cite{Micklitz2017_PRL, Sumita2018}, are characterized by a zero-dimensional (0D) index~\cite{Kobayashi2018}.
In addition, for even-parity superconductivity, such line nodes are also protected by a one-dimensional (1D) winding number, which is robust even when the mirror (glide) symmetry is broken~\cite{Kobayashi2018}.
As just described, topological property helps us to easily understand the stability of nodes.
Therefore it is desired to complementarily classify superconducting gap structures using both group theory and topological theory.

On the other hand, nodes on high-symmetry lines have not been fully clarified.
In particular for $j_z$-dependent gap structures~\cite{Nomoto2016_PRB, Sumita2018}, the topological protection of nodes has not been investigated.
Thus, in this paper, we classify irreducible representations (IRs) of any $\bm{k}$ points on high-symmetry $n$-fold ($n = 2$, $3$, $4$, and $6$) lines, from the viewpoints of both group theory and topological protection.
This results in the topological classification of the superconducting gap structure on the line, which includes the $j_z$-dependent gap structures, assuming that the line intersects a normal-state Fermi surface (FS).
Such classification clarifies four types of the gap structure: full gap, point nodes, line nodes, and surface nodes (Bogoliubov Fermi surface).
As a result, it is shown that the classification by group theory completely agrees with the topological classification.
All nodes on the high-symmetry line are characterized by a 0D $\mathbb{Z}$ or $\mathbb{Z}_2$ index, which is defined using the crystal symmetry, and therefore, we call such nodes \textit{topological crystalline superconducting nodes}.
Especially, \textit{topological crystalline point nodes} are a novel type of nodes, which is different from well-known Weyl nodes characterized by a two-dimensional (2D) Chern number~\cite{Daido2016, Yanase2016, Yanase2017, Kozii2016, Venderbos2016, Sau2012, Goswami2015, Fischer2014, Meng2012, Yang2014, Volovik2017}.

This paper is constructed as follows.
First, in Sec.~\ref{sec:preparation}, we make a remark about some terminologies and notations, which are used throughout the paper, for the avoidance of confusion.
In Sec.~\ref{sec:gap_classification_group_theory}, we review the results of the group-theoretical analysis of the superconducting gap on the $n$-fold axes in the BZ~\cite{Sumita2018}.
Next, we examine the topological protection of nodes on the high-symmetry line by using the Wigner criterion (Herring test) and the orthogonality test in Sec.~\ref{sec:gap_classification_topological}.
We also give an intuitive understanding of the tests by showing simple examples (Sec.~\ref{sec:gap_classification_topological_example}).
Furthermore, in Sec~\ref{sec:candidates}, we suggest the topological crystalline superconducting nodes in some candidate superconductors: SrPtAs (Sec.~\ref{sec:SrPtAs}), CeCoIn$_5$ (Sec.~\ref{sec:CeCoIn5}), UPt$_3$ (Sec.~\ref{sec:UPt3}), and UCoGe (Sec.~\ref{sec:UCoGe}).
Finally, a brief summary and discussion are given in Sec.~\ref{sec:summary}.

\section{Preparation}
\label{sec:preparation}
In this section, we define some terminologies and notations which are commonly used in the group-theoretical classification (Sec.~\ref{sec:gap_classification_group_theory}) and the topological classification (Sec.~\ref{sec:gap_classification_topological}).
In all the discussions below, we assume that the system is \textit{centrosymmetric} and \textit{paramagnetic}, for comparison with the previous result in Ref.~\cite{Sumita2018}.\footnote{Indeed, our new classification theory (Sec.~\ref{sec:gap_classification_topological}) can also be applied to other cases, e.g. noncentrosymmetric superconductors.}

First, we focus on a magnetic space group $M$; according to the above assumptions, $M$ is equal to $G + {\cal T} G$, where $G$ is a unitary part of $M$ including the spacial inversion ${\cal I} = \{I | \bm{0}\}$, and ${\cal T} = \{T | \bm{0}\}$ is the time-reversal symmetry (TRS).\footnote{Note that the notation $\{p | \bm{a}\}$ is a conventional Seitz space group symbol with a point-group operation $p$ and a translation $\bm{a}$.}
In order to classify the gap structure on high-symmetry $n$-fold ($n = 2$, $3$, $4$, and $6$) axes in the BZ, we restrict $G$ ($M$) to ${\cal G}^{\bm{k}} \subset G$ (${\cal M}^{\bm{k}} \subset M$), which is the (magnetic) little group leaving $\bm{k}$ points on the axes invariant modulo a reciprocal lattice vector.
Thus the factor group of the (magnetic) little group by the translation group ${\cal G}^{\bm{k}} / \mathbb{T}$ (${\cal M}^{\bm{k}} / \mathbb{T}$), which is called the (magnetic) little cogroup, is an $n$-fold rotational-symmetric group $\bar{\cal G}^{\bm{k}} = C_n$ or $C_{nv}$ ($\bar{\cal M}^{\bm{k}} = C_n + {\cal T I} C_n$ or $C_{nv} + {\cal T I} C_{nv}$).
Here, $\{z_{g, h}^{\bm{k}}\} \in Z^2(M / \mathbb{T}, U(1)_\phi)$ is a factor system arising in a representation on $\bm{k}$,
\begin{equation}
 z_{g, h}^{g h \bm{k}} U^{\bm{k}}(g h) =
  \begin{cases}
   U^{h \bm{k}}(g) U^{\bm{k}}(h) & \phi(g) = 1, \\
   U^{h \bm{k}}(g) U^{\bm{k}}(h)^* & \phi(g) = -1,
  \end{cases}
\end{equation}
where $\phi: \bar{\cal M}^{\bm{k}} \to \mathbb{Z}_2 = \{\pm 1\}$ is an indicator for unitary/antiunitary symmetry.

Next, we define $\lambda^{\bm{k}}_\alpha(m)$ as a double-valued small corepresentation of symmetry operations $m \in {\cal M}^{\bm{k}}$, which represents the normal Bloch state with the crystal momentum $\bm{k}$.
$\alpha$ is the label of the double-valued IR, which corresponds to the total angular momentum of the Bloch state $j_z = \pm 1 / 2, \pm 3 / 2, \dotsc$ in spin-orbit coupled systems.\footnote{Strictly speaking, $\alpha$ is an double-valued IR of the finite group $\bar{\cal M}^{\bm{k}}$ while $j_z$ is a basis of the continuous group. Therefore there is no one-to-one correspondence between $\alpha$ and $j_z$ in some cases. In the $C_{2v}$ symmetry, for example, the IR $\alpha = 1 / 2$ includes all normal Bloch states with half-integer total angular momentum $j_z = \pm 1 / 2, \pm 3 / 2, \pm 5 / 2, \dotsc$. In this paper, however, we represent $j_z$ with minimum absolute value, which satisfies $j_z \downarrow \bar{\cal M}^{\bm{k}} = \alpha$, as the angular-momentum counterpart of the IR $\alpha$.}
Since ${\cal M}^{\bm{k}}$ is the semi-direct product between the magnetic little cogroup $\bar{\cal M}^{\bm{k}}$ and the translation group $\mathbb{T}$, $\lambda^{\bm{k}}_\alpha(m)$ is equal to $\bar{\lambda}^{\bm{k}}_\alpha(\bar{m}) F^{\bm{k}}(t)$, where $\bar{\lambda}^{\bm{k}}_\alpha$ ($F^{\bm{k}}$) is the IR of $\bar{\cal M}^{\bm{k}}$ ($\mathbb{T}$), and $m = \bar{m}t$ for $\bar{m} \in \bar{\cal M}^{\bm{k}}$ and $t \in \mathbb{T}$.
$\bar{\lambda}^{\bm{k}}_\alpha$ is constructed from the projective IR $\bar{\gamma}^{\bm{k}}_\alpha$ of the (unitary) little cogroup $\bar{\cal G}^{\bm{k}}$ with the appropriate factor system $\{z_{g, h}^{\bm{k}}\}$~\cite{Bradley1968, Bradley}, by using the Wigner criterion (Herring test)~\cite{Wigner, Herring1937, Inui-Tanabe-Onodera, Bradley, Shiozaki2018_arXiv}:
\begin{equation}
 W_\alpha^{\mathfrak{T}} \equiv \frac{1}{|\bar{\cal G}^{\bm{k}}|} \sum_{\bar{g} \in \bar{\cal G}^{\bm{k}}} z_{\mathfrak{T} \bar{g}, \mathfrak{T} \bar{g}}^{\bm{k}} \chi[\bar{\gamma}^{\bm{k}}_\alpha((\mathfrak{T} \bar{g})^2)] =
 \begin{cases}
  1 & \text{(a)}, \\
  -1 & \text{(b)}, \\
  0 & \text{(c)},
 \end{cases}
\end{equation}
where $\chi$ is the character of the representation, $\mathfrak{T} \equiv {\cal T I}$ is a TRS-like operator preserving $\bm{k}$.
In the (b) and (c) cases, the degeneracy of $\bar{\lambda}^{\bm{k}}_\alpha$ is twice as much as that of $\bar{\gamma}^{\bm{k}}_\alpha$, while $\bar{\lambda}^{\bm{k}}_\alpha(\bar{g})$ gives the same representation as $\bar{\gamma}^{\bm{k}}_\alpha(\bar{g})$ for $\bar{g} \in \bar{\cal G}^{\bm{k}}$ in the (a) case (for details, see Appendix~\ref{sec:Wigner_criterion}).
In the above discussion, we have introduced a lot of notations for groups and representations.
For the avoidance of confusion, we summarize the intriguing notations in Table~\ref{tab:notations}, and show a simple example in the next section.

\begin{table*}[tbp]
 \caption{Notations of some intriguing groups and representations. The terminologies in the first column are associated with an unitary group, while those in the third column are with a nonunitary group including antiunitary operators. In the table, we adopt the terminologies of Ref.~\cite{Bradley}.}
 \label{tab:notations}
 \begin{center}
  \begin{tabular}{lcp{5mm}lcp{10mm}l} \hline\hline
   \multicolumn{1}{c}{Terminology} & Notation & & \multicolumn{1}{c}{Terminology} & Notation & & \multicolumn{1}{c}{Definition} \\ \hline
   Space group          & $G$                            & & Magnetic space group    & $M$                             & & Whole crystal symmetry of the system   \\
   Little group         & ${\cal G}^{\bm{k}}$            & & Magnetic little group   & ${\cal M}^{\bm{k}}$             & & Stabilizer of $\bm{k}$                 \\
   Small representation & $\gamma^{\bm{k}}_\alpha$       & & Small corepresentation  & $\lambda^{\bm{k}}_\alpha$       & & IR of (magnetic) little group          \\
   Little cogroup       & $\bar{\cal G}^{\bm{k}}$        & & Magnetic little cogroup & $\bar{\cal M}^{\bm{k}}$         & & Factor group of (magnetic) little group by $\mathbb{T}$ \\
   N/A                  & $\bar{\gamma}^{\bm{k}}_\alpha$ & & N/A                     & $\bar{\lambda}^{\bm{k}}_\alpha$ & & IR of (magnetic) little cogroup        \\ \hline\hline
  \end{tabular}
 \end{center}
\end{table*}

\subsection{Example}
We consider a unitary space group $G = P2/m$ as a simple example, which helps us understand the terminologies in Table~\ref{tab:notations}.
Assuming the paramagnetic system, we consider a magnetic space group $M = G + {\cal T} G = P2/m1'$.
Now we focus on a twofold ($C_2$-symmetric) axis $\Gamma$-$Z$ in the BZ, where the (magnetic) little group is
\begin{align}
 {\cal G}^{\bm{k}} &= \mathbb{T} + C_2 \mathbb{T}, \\
 {\cal M}^{\bm{k}} &= {\cal G}^{\bm{k}} + \mathfrak{T} {\cal G}^{\bm{k}} = \mathbb{T} + C_2 \mathbb{T} + \mathfrak{T} \mathbb{T} + \mathfrak{T} C_2 \mathbb{T}.
\end{align}
Therefore the (magnetic) little cogroup is
\begin{align}
 \bar{\cal G}^{\bm{k}} &= {\cal G}^{\bm{k}} / \mathbb{T} = \{E, C_2\}, \\
 \bar{\cal M}^{\bm{k}} &= {\cal M}^{\bm{k}} / \mathbb{T} = \{E, C_2, \mathfrak{T}, \mathfrak{T} C_2\}.
\end{align}
Taking half-integer spin states into account, we find that the factor system is
\begin{equation}
 (C_2)^2 = - E, \, \mathfrak{T}^2 = - E, \, \mathfrak{T} C_2 = C_2 \mathfrak{T}.
\end{equation}
Note that in the above equation, we treat the symbols ($E$, $C_2$, and $\mathfrak{T}$) as not elements of the group but representation matrices operating the Hamiltonian.
A similar treatment is applied to other equations representing commutation relations.

The projective IRs $\bar{\gamma}^{\bm{k}}_{\pm 1 / 2}$ of the little cogroup $\bar{\cal G}^{\bm{k}}$, with the factor system $(C_2)^2 = - E$, are given by
\begin{equation}
 \bar{\gamma}^{\bm{k}}_{\pm 1 / 2}(E) = 1, \, \bar{\gamma}^{\bm{k}}_{\pm 1 / 2}(C_2) = \pm i, \label{eq:Bloch_rep_C2}
\end{equation}
which corresponds to spin-up and spin-down states, respectively.
For the spin-up one ($\alpha = + 1 / 2$), the Wigner criterion~\cite{Wigner, Herring1937, Inui-Tanabe-Onodera, Bradley, Shiozaki2018_arXiv} for TRS $\mathfrak{T}$ is
\begin{align}
 W_{+ 1 / 2}^{\mathfrak{T}} &= \frac{1}{2} \sum_{\bar{g} \in \{E, C_2\}} z_{\mathfrak{T} \bar{g}, \mathfrak{T} \bar{g}}^{\bm{k}} \chi[\bar{\gamma}^{\bm{k}}_{+ 1 / 2}((\mathfrak{T} \bar{g})^2)] \notag \\
 &= \frac{1}{2} \{ - 1 + 1 \} = 0,
\end{align}
which indicates that the spin-up state ($\alpha = + 1 / 2$) is degenerated with the spin-down state ($\alpha = - 1 / 2$).
Thus the representation $\bar{\lambda}^{\bm{k}}$ of the magnetic little cogroup $\bar{\cal M}^{\bm{k}}$ can be constructed as follows:
\begin{equation}
 \begin{array}{ccccc} \hline\hline
  \bar{m} & E & C_2 & \mathfrak{T} & \mathfrak{T} C_2 \\ \hline
  \bar{\lambda}^{\bm{k}}(\bar{m}) & \begin{pmatrix} 1 & 0 \\ 0 & 1 \end{pmatrix} & \begin{pmatrix} + i & 0 \\ 0 & - i \end{pmatrix} & \begin{pmatrix} 0 & 1 \\ - 1 & 0 \end{pmatrix} & \begin{pmatrix} 0 & i \\ i & 0 \end{pmatrix} \\ \hline\hline
 \end{array}
\end{equation}
By using the above representation, the small corepresentation of $m \in {\cal M}^{\bm{k}}$ is given by
\begin{equation}
 \lambda^{\bm{k}}(m) = \bar{\lambda}^{\bm{k}}(\bar{m}) F^{\bm{k}}(t) = \bar{\lambda}^{\bm{k}}(\bar{m}) e^{- i \bm{k} \cdot \bm{t}},
\end{equation}
where $m = \bar{m} t$ with $\bar{m} \in \bar{\cal M}^{\bm{k}}$ and $t = \{E | \bm{t}\} \in \mathbb{T}$.
This simple example of the $C_2$ axis is also used in the later discussion (Sec.~\ref{sec:gap_classification_topological_example}).

\section{Group-theoretical classification of superconducting gap}
\label{sec:gap_classification_group_theory}
In this section, we review the group-theoretical classification of the superconducting gap on the $n$-fold axes~\cite{Sumita2018}.

\subsection{Classification theory}
Let us briefly introduce the superconducting gap classification in terms of the group theory.
As seen in Sec.~\ref{sec:preparation}, we can obtain the magnetic small representation $\lambda^{\bm{k}}_\alpha$ corresponding to the normal Bloch state on a high-symmetry $\bm{k}$ point.
In the superconducting state, zero center-of-mass momentum Cooper pairs have to be formed between degenerate states present at $\bm{k}$ and $- \bm{k}$ in the same band when we adopt the weak-coupling BCS theory.
Then, the two states should be connected by spacial inversion ${\cal I}$ and/or TRS ${\cal T}$, except for an accidentally degenerate case.
As a result, the representation $P^{\bm{k}}_\alpha$ of the Cooper-pair wave function can be constructed from the representations of the Bloch state $\lambda^{\bm{k}}_\alpha$.

Next, we calculate the single-valued representation $P^{\bm{k}}_\alpha$ of the Cooper-pair wave function.
Taking into account the antisymmetry of the Cooper pairs and the degeneracy of the two states, we can regard $P^{\bm{k}}_\alpha$ as an antisymmetrized Kronecker square~\cite{Bradley, Bradley1970}, with zero total momentum, of the induced representation $\lambda^{\bm{k}}_\alpha \uparrow {\cal M}^{\bm{k}}_{\text{pair}}$.
Here, ${\cal M}^{\bm{k}}_{\text{pair}} \equiv {\cal M}^{\bm{k}} + {\cal I M}^{\bm{k}}$ is the group to which the representation $P^{\bm{k}}_\alpha$ of the pair wave function belongs.
This is given by $P^{\bm{k}}_\alpha(m) = \bar{P}^{\bm{k}}_\alpha(\bar{m})$, where $m = \bar{m}t$ for $m \in {\cal M}^{\bm{k}}_{\text{pair}}$, $\bar{m} \in \bar{\cal M}^{\bm{k}}_{\text{pair}} \equiv {\cal M}^{\bm{k}}_{\text{pair}} / \mathbb{T}$, and $t \in \mathbb{T}$.
$\bar{P}^{\bm{k}}_\alpha$, which is the representation of $\bar{\cal M}^{\bm{k}}_{\text{pair}}$, is obtained in a systematic way by using the double coset decomposition and the corresponding Mackey-Bradley theorem~\cite{Bradley, Bradley1970, Mackey1953},
\begin{subequations}
 \label{eq:Mackey-Bradley}
 \begin{align}
  \chi[\bar{P}^{\bm{k}}_\alpha(\bar{m})] &= \frac{z_{\bar{m}, {\cal I}}^{\bm{k}}}{z_{{\cal I}, {\cal I} \bar{m} {\cal I}}^{\bm{k}}} \chi[\bar{\lambda}^{\bm{k}}_\alpha(\bar{m})] \chi[\bar{\lambda}^{\bm{k}}_\alpha({\cal I} \bar{m} {\cal I})], \label{eq:Mackey-Bradley_a} \\
  \chi[\bar{P}^{\bm{k}}_\alpha({\cal I} \bar{m})] &= - \frac{z_{\bar{m}, {\cal I}}^{\bm{k}} z_{\bar{m} {\cal I}, \bar{m}}^{\bm{k}}}{z_{{\cal I}, {\cal I} \bar{m} {\cal I} \bar{m}}^{\bm{k}}} \chi[\bar{\lambda}^{\bm{k}}_\alpha({\cal I} \bar{m} {\cal I} \bar{m})], \label{eq:Mackey-Bradley_b}
 \end{align}
\end{subequations}
where $\bar{m} \in \bar{\cal M}^{\bm{k}}$.

Finally, we reduce $\bar{P}^{\bm{k}}_\alpha$ into single-valued IRs of the point group $\bar{\cal M}^{\bm{k}}_{\text{pair}}$.
The gap functions should be zero, and thus, the gap nodes appear, if the corresponding IRs do not exist in the results of reductions~\cite{Izyumov1989, Yarzhemsky1992, Yarzhemsky1998}.
Otherwise, the superconducting gap will open in general.
Therefore the representation $\bar{P}^{\bm{k}}_\alpha$ of pair wave function tells us the presence or absence of superconducting gap nodes.

\subsection{Classification results on high-symmetry lines}
Now we classify the superconducting gap on high-symmetry lines.
We apply the above group-theoretical analysis to $n$-fold axes in the BZ, where the magnetic little cogroup is $\bar{\cal M}^{\bm{k}} = C_{n(v)} + \mathfrak{T} C_{n(v)}$.
The results of classification are summarized in Table~\ref{tab:gap_classification_group_theory}.

\begin{table}[tbp]
 \caption{Group-theoretical classification of superconducting gap on high-symmetry lines where the little cogroup is $\bar{\cal G}^{\bm{k}}$. $\bar{P}^{\bm{k}}_\alpha$ on each line is decomposed by IRs of corresponding $\bar{\cal G}^{\bm{k}}_{\text{pair}} = \bar{\cal G}^{\bm{k}} + {\cal I} \bar{\cal G}^{\bm{k}}$. The labels of IR ($\alpha$) are represented by the subscripts $1 / 2$, $3 / 2$, and $5 / 2$.}
 \label{tab:gap_classification_group_theory}
 \begin{center}
  \begin{tabular}{lll} \hline\hline
   $\bar{\cal G}^{\bm{k}}$ ($\bar{\cal G}^{\bm{k}}_{\text{pair}}$) & $\bar{\lambda}^{\bm{k}}_\alpha$ & $\bar{P}^{\bm{k}}_\alpha$ \\ \hline
   $C_2$ ($C_{2h}$)    & $E_{1 / 2}$            & $A_g + A_u + 2 B_u$                 \\
   $C_3$ ($S_6$)       & $E_{1 / 2}$            & $A_g + A_u + E_u$                   \\
                       & $2 B_{3 / 2}$          & $A_g + 3 A_u$                       \\
   $C_4$ ($C_{4h}$)    & $E_{1 / 2}, E_{3 / 2}$ & $A_g + A_u + E_u$                   \\
   $C_6$ ($C_{6h}$)    & $E_{1 / 2}, E_{5 / 2}$ & $A_g + A_u + E_u$                   \\
                       & $E_{3 / 2}$            & $A_g + A_u + 2 B_u$                 \\
   $C_{2v}$ ($D_{2h}$) & $E_{1 / 2}$            & $A_g + A_u + B_{2u} + B_{3u}$       \\
   $C_{3v}$ ($D_{3d}$) & $E_{1 / 2}$            & $A_{1g} + A_{1u} + E_u$             \\
                       & $E_{3 / 2}$            & $A_{1g} + 2 A_{1u} + A_{2u}$        \\
   $C_{4v}$ ($D_{4h}$) & $E_{1 / 2}, E_{3 / 2}$ & $A_{1g} + A_{1u} + E_u$             \\
   $C_{6v}$ ($D_{6h}$) & $E_{1 / 2}, E_{5 / 2}$ & $A_{1g} + A_{1u} + E_{1u}$          \\
                       & $E_{3 / 2}$            & $A_{1g} + A_{1u} + B_{1u} + B_{2u}$ \\ \hline\hline
  \end{tabular}
 \end{center}
\end{table}

As shown in Table~\ref{tab:gap_classification_group_theory}, the pair wave function on the $C_{2(v)}$- and $C_{4(v)}$-symmetric lines has a unique representation irrespective of the IR $\alpha$, namely the total angular momentum of the normal Bloch state.
On the $C_{3(v)}$- and $C_{6(v)}$-symmetric axes, on the other hand, there are two nonequivalent representations of the pair wave function depending on the angular momentum $j_z$.
For example, the $E_u$-symmetric gap function on the $C_{3(v)}$ axis fully opens the gap if the total angular momentum $j_z$ of the normal Bloch state is $\pm 1 / 2$, while the function creates point nodes if $j_z = \pm 3 / 2$.
In Ref.~\cite{Sumita2018}, we called it \textit{$j_z$-dependent point nodes}.
Such nodes have been already suggested in some candidate superconductors~\cite{Yarzhemsky1992, Yarzhemsky1998, Nomoto2016_PRL, Sumita2018, Nomoto2016_PRB}.

\section{Topological protection of nodes on high-symmetry lines}
\label{sec:gap_classification_topological}
In this section, we consider the topological protection of nodes on high-symmetry lines summarized in Table~\ref{tab:gap_classification_group_theory}.

\subsection{Method}
First, we introduce the method for topological classification of the nodes.
As mentioned in Sec.~\ref{sec:preparation}, a unitary little cogroup on the $C_{n(v)}$-symmetric line in the BZ is denoted by $\bar{\cal G}^{\bm{k}}$, and a IR of $\bar{\cal G}^{\bm{k}}$ by $\alpha$, which corresponds to the total angular momentum of the normal Bloch state $j_z = \pm 1/2, \pm 3/2, \dotsc$.
Here, we define TRS, particle-hole symmetry (PHS), and chiral symmetry (CS) like operators preserving any $\bm{k}$ points by $\mathfrak{T} \equiv {\cal T I}$, $\mathfrak{C} \equiv {\cal C I}$, and $\Gamma \equiv {\cal T C}$, respectively.
Thus the intriguing symmetry is represented by the following group:
\begin{equation}
 \bar{\mathfrak{G}}^{\bm{k}} = \bar{\cal G}^{\bm{k}} + \mathfrak{T} \bar{\cal G}^{\bm{k}} + \mathfrak{C} \bar{\cal G}^{\bm{k}} + \Gamma \bar{\cal G}^{\bm{k}}.
\end{equation}
Then, using the factor system $\{z_{g, h}^{\bm{k}}\} \in Z^2(\bar{\mathfrak{G}}^{\bm{k}}, \mathrm{U}(1)_\phi)$, we execute the \textit{Wigner criteria}~\cite{Wigner, Herring1937, Inui-Tanabe-Onodera, Bradley, Shiozaki2018_arXiv} for $\mathfrak{T}$ and $\mathfrak{C}$,
\begin{align}
 W_\alpha^{\mathfrak{T}} &\equiv \frac{1}{|\bar{\cal G}^{\bm{k}}|} \sum_{g \in \bar{\cal G}^{\bm{k}}} z_{\mathfrak{T} g, \mathfrak{T} g}^{\bm{k}} \chi[\bar{\gamma}^{\bm{k}}_\alpha((\mathfrak{T} g)^2)] =
 \begin{cases}
  1, \\
  -1, \\
  0,
 \end{cases} \label{eq:Wigner_criterion_T} \\
 W_\alpha^{\mathfrak{C}} &\equiv \frac{1}{|\bar{\cal G}^{\bm{k}}|} \sum_{g \in \bar{\cal G}^{\bm{k}}} z_{\mathfrak{C} g, \mathfrak{C} g}^{\bm{k}} \chi[\bar{\gamma}^{\bm{k}}_\alpha((\mathfrak{C} g)^2)] =
 \begin{cases}
  1, \\
  -1, \\
  0,
 \end{cases} \label{eq:Wigner_criterion_C}
 \intertext{and the \textit{orthogonality test}~\cite{Inui-Tanabe-Onodera, Shiozaki2018_arXiv} for $\Gamma$:}
 W_\alpha^\Gamma &\equiv \frac{1}{|\bar{\cal G}^{\bm{k}}|} \sum_{g \in \bar{\cal G}^{\bm{k}}} \frac{z_{g, \Gamma}^{\bm{k} *}}{z_{\Gamma, \Gamma^{-1} g \Gamma}^{\bm{k} *}} \chi[\bar{\gamma}^{\bm{k}}_\alpha(\Gamma^{-1} g \Gamma)^*] \chi[\bar{\gamma}^{\bm{k}}_\alpha(g)] =
 \begin{cases}
  1, \\
  0.
 \end{cases} \label{eq:orthogonality_test_G}
\end{align}
In the above tests, we investigate the orthogonality between $\{c_{\bm{k}, \alpha, i}^\dagger\}$ and $\{a c_{\bm{k}, \alpha, i}^\dagger a^{-1}\}$ ($a = \mathfrak{T}$, $\mathfrak{C}$, or $\Gamma$), where $c_{\bm{k}, \alpha, i}^\dagger$ is the $i$-th basis of the IR $\bar{\gamma}^{\bm{k}}_\alpha$ (for details, see Appendixes~\ref{sec:Wigner_criterion} and \ref{sec:orthogonality_test}).
From Eqs.~\eqref{eq:Wigner_criterion_T}-\eqref{eq:orthogonality_test_G}, we obtain the set of $(W_\alpha^{\mathfrak{T}}, W_\alpha^{\mathfrak{C}}, W_\alpha^\Gamma)$, which indicates the \textit{emergent} Altland-Zirnbauer (EAZ) symmetry class of the Bogoliubov-de Gennes (BdG) Hamiltonian on the high-symmetry line by using the knowledge of $K$ theory~\cite{Altland1997, Shiozaki2018_arXiv}.
Table~\ref{tab:emergent_AZ_class} shows the correspondence between the set of $(W_\alpha^{\mathfrak{T}}, W_\alpha^{\mathfrak{C}}, W_\alpha^\Gamma)$ and the EAZ symmetry class.

\begin{table}
 \caption{Correspondence table between the set of $(W_\alpha^{\mathfrak{T}}, W_\alpha^{\mathfrak{C}}, W_\alpha^\Gamma)$ and the EAZ symmetry class. The fifth column shows a topological classification of the IR at the $\bm{k}$ point for each EAZ class.}
 \label{tab:emergent_AZ_class}
 \begin{center}
  \begin{tabular}{ccccc} \hline\hline
   $W_\alpha^{\mathfrak{T}}$ & $W_\alpha^{\mathfrak{C}}$ & $W_\alpha^\Gamma$ & EAZ class & Classification \\ \hline
   $0$  & $0$  & $0$ & A    & $\mathbb{Z}$   \\
   $0$  & $0$  & $1$ & AIII & $0$            \\
   $+1$ & $0$  & $0$ & AI   & $\mathbb{Z}$   \\
   $+1$ & $+1$ & $1$ & BDI  & $\mathbb{Z}_2$ \\
   $0$  & $+1$ & $0$ & D    & $\mathbb{Z}_2$ \\
   $-1$ & $+1$ & $1$ & DIII & $0$            \\
   $-1$ & $0$  & $0$ & AII  & $2\mathbb{Z}$  \\
   $-1$ & $-1$ & $1$ & CII  & $0$            \\
   $0$  & $-1$ & $0$ & C    & $0$            \\
   $+1$ & $-1$ & $1$ & CI   & $0$            \\ \hline\hline
  \end{tabular}
 \end{center}
\end{table}

Furthermore, from the EAZ symmetry class, we can classify the IR at the $\bm{k}$ point into $0$, $\mathbb{Z}$, $2\mathbb{Z}$ or $\mathbb{Z}_2$ (Table~\ref{tab:emergent_AZ_class}).
In this context, $(W_\alpha^{\mathfrak{T}}, W_\alpha^{\mathfrak{C}}, W_\alpha^\Gamma)$ gives a symmetry-based topological classification of the Hamiltonian at each $\bm{k}$ point on the line.
Therefore, when the line intersects a normal-state FS, the above information is nothing but a topological classification of superconducting gap nodes on the line; when the classification is nontrivial ($\mathbb{Z}$, $2\mathbb{Z}$, or $\mathbb{Z}_2$), the intersection leads to a node characterized by the topological invariant. Otherwise, a gap opens at the intersection point.

\subsection{Examples}
\label{sec:gap_classification_topological_example}
In the previous section, we explained the systematic method for the topological classification of superconducting gap nodes on a high-symmetry line, by using the EAZ symmetry class.
In this section, we see the meaning of the classification by taking a gap and a node on a $C_2$-symmetric line as examples.

\subsubsection{Gap on a \texorpdfstring{$C_2$}{C2} line}
Now we consider the $A_g$ pair wave function on a $C_2$-symmetric line in the BZ.
According to the group-theoretical gap classification, this function opens a superconducting gap on the line (see Table~\ref{tab:gap_classification_group_theory}).
We here confirm the result in terms of topological classification based on the EAZ symmetry class.

In this case, the little cogroup is
\begin{equation}
 \bar{\cal G}^{\bm{k}} = \bigl\{ \{E | \bm{0}\}, \{C_2 | \bm{0}\} \bigr\}. \label{eq:little_cogroup_C2}
\end{equation}
Taking into account $(C_2)^2 = - E$, the IR matrices of $\bar{\cal G}^{\bm{k}}$ are given by Eq.~\eqref{eq:Bloch_rep_C2}.
Here we choose the representation matrix $\bar{\gamma}^{\bm{k}}_{+ 1 / 2}$ of the IR $\alpha = + 1 / 2$ and apply the above topological classification.\footnote{In this case, the final result is not changed even if we choose the other IR $\bar{\gamma}^{\bm{k}}_{- 1 / 2}$.}
Before going to the tests for the EAZ symmetry class, we prepare some relationships among the intriguing operators.
Assuming the symmorphic system, we obtain the relationships about $\mathfrak{T}$:
\begin{equation}
 \mathfrak{T}^2 = - E, \, [\mathfrak{T}, C_2] = 0. \label{eq:commutation_C2_Ag_T}
\end{equation}
Furthermore, for the $A_g$ pairing, we get
\begin{equation}
 \mathfrak{C}^2 = + E, \, [\mathfrak{C}, C_2] = 0. \label{eq:commutation_C2_Ag_C}
\end{equation}
Therefore
\begin{equation}
 [\Gamma, C_2] = [\mathfrak{T}, C_2] \mathfrak{C} + \mathfrak{T} [\mathfrak{C}, C_2] = 0, \label{eq:commutation_C2_Ag_G}
\end{equation}
where we use $\Gamma = {\cal T C} = \mathfrak{T C}$.

Now we apply the topological classification to the normal Bloch state $\bar{\gamma}^{\bm{k}}_{+ 1 / 2}$ and the $A_g$ pair wave function.
Using the above relationships, the Wigner criteria for $\mathfrak{T}$ and $\mathfrak{C}$ and the orthogonality test for $\Gamma$ are given by
\begin{align}
 W_{+ 1 / 2}^{\mathfrak{T}} &= \frac{1}{2} \left\{ z_{\mathfrak{T}, \mathfrak{T}}^{\bm{k}} \chi[\bar{\gamma}^{\bm{k}}_{+ 1 / 2}(\mathfrak{T}^2)] \right. \notag \\
 &\quad \left. + z_{\mathfrak{T} C_2, \mathfrak{T} C_2}^{\bm{k}} \chi[\bar{\gamma}^{\bm{k}}_{+ 1 / 2}((\mathfrak{T} C_2)^2)] \right\} \notag \\
 &= \frac{1}{2} \{ - 1 + 1 \} = 0, \label{eq:Wigner_criterion_C2_Ag_T} \\
 W_{+ 1 / 2}^{\mathfrak{C}} &= \frac{1}{2} \left\{ z_{\mathfrak{C}, \mathfrak{C}}^{\bm{k}} \chi[\bar{\gamma}^{\bm{k}}_{+ 1 / 2}(\mathfrak{C}^2)] \right. \notag \\
 &\quad \left. + z_{\mathfrak{C} C_2, \mathfrak{C} C_2}^{\bm{k}} \chi[\bar{\gamma}^{\bm{k}}_{+ 1 / 2}((\mathfrak{C} C_2)^2)] \right\} \notag \\
 &= \frac{1}{2} \{ + 1 - 1 \} = 0, \label{eq:Wigner_criterion_C2_Ag_C} \\
 W_{+ 1 / 2}^\Gamma &= \frac{1}{2} \left\{ \frac{z_{E, \Gamma}^{\bm{k}}}{z_{\Gamma, \Gamma^{-1} E \Gamma}^{\bm{k}}} \chi[\bar{\gamma}^{\bm{k}}_{+ 1 / 2}(E)^*] \chi[\bar{\gamma}^{\bm{k}}_{+ 1 / 2}(E)] \right. \notag \\
 &\quad \left. + \frac{z_{C_2, \Gamma}^{\bm{k} *}}{z_{\Gamma, \Gamma^{-1} C_2 \Gamma}^{\bm{k} *}} \chi[\bar{\gamma}^{\bm{k}}_{+ 1 / 2}(\Gamma^{-1} C_2 \Gamma)^*] \chi[\bar{\gamma}^{\bm{k}}_{+ 1 / 2}(C_2)] \right\} \notag \\
 &= \frac{1}{2} \{ 1 + 1 \} = 1. \label{eq:orthogonality_test_C2_Ag_G}
\end{align}
Thus, according to Table~\ref{tab:emergent_AZ_class}, the system is identified as the EAZ symmetry class AIII.
Since the class AIII is classified into $0$, the gap classification is topologically trivial.
This means that the $A_g$ pair wave function opens a gap on the $C_2$-symmetric axis, which is consistent with the group-theoretical classification.

\begin{figure}[tbp]
 \centering
 \includegraphics[width=8cm, clip]{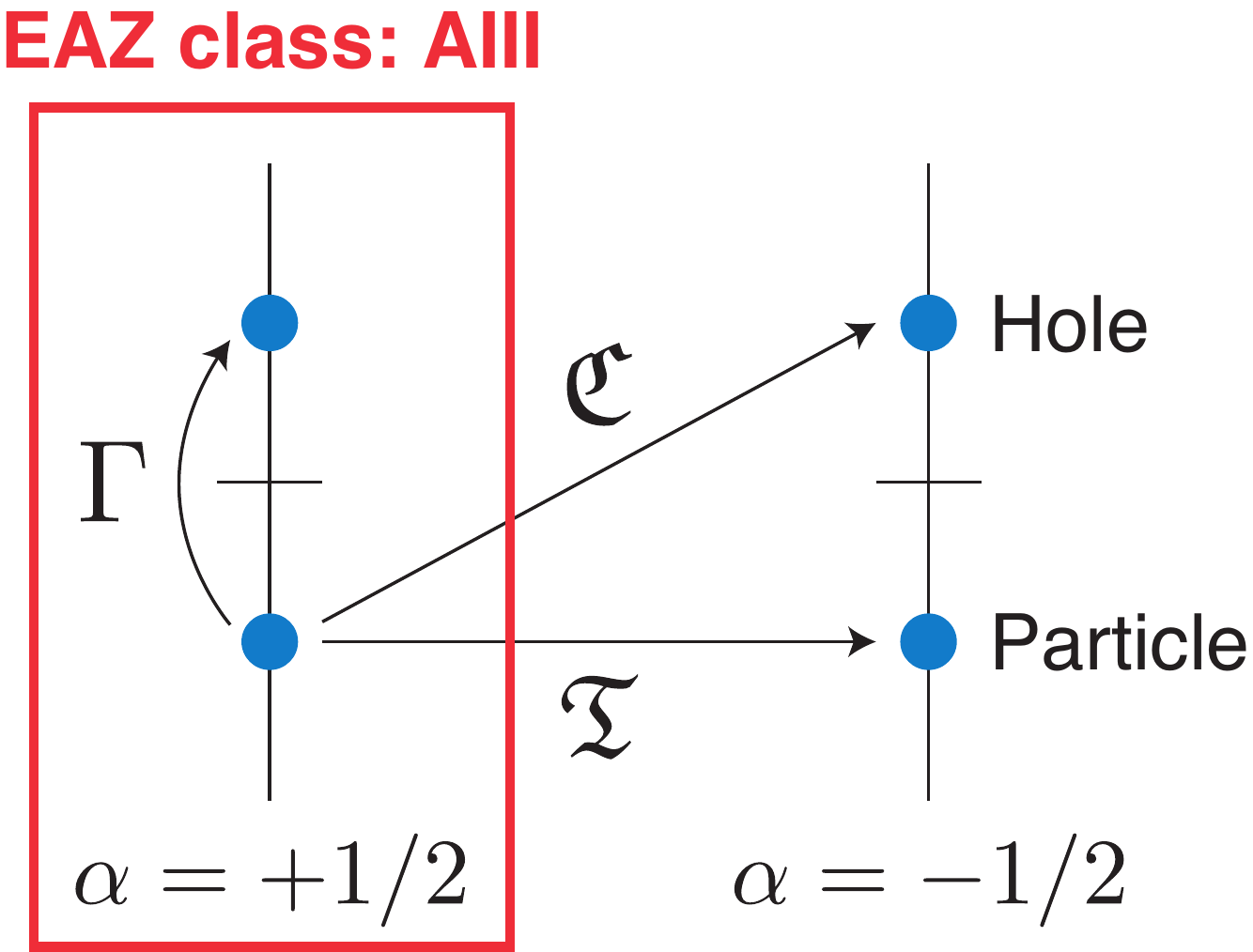}
 \caption{Schematic picture of the BdG Hamiltonian with the $A_g$ pair wave function on a $C_2$-symmetric line. The space in the red frame belongs to the EAZ class AIII.}
 \label{fig:emergent_AZ_class_C2_Ag}
\end{figure}

Here we explain the meaning of the EAZ symmetry class.
Figure~\ref{fig:emergent_AZ_class_C2_Ag} represents a schematic picture of the BdG Hamiltonian with the $A_g$ pair wave function.
In the above discussion, we started from the representation matrix $\bar{\gamma}^{\bm{k}}_{+ 1 / 2}$ of the IR $\alpha = + 1 / 2$, which corresponds to the normal Bloch state (the lower left particle in Fig.~\ref{fig:emergent_AZ_class_C2_Ag}).
The Wigner criterion for the TRS-like operator $\mathfrak{T}$ [Eq.~\eqref{eq:Wigner_criterion_C2_Ag_T}] results in $W_{+ 1 / 2}^{\mathfrak{T}} = 0$, which indicates that $\mathfrak{T}$ gives a basis of the nonequivalent IR (see Appendix~\ref{sec:Wigner_criterion}).
Therefore the lower left particle in Fig.~\ref{fig:emergent_AZ_class_C2_Ag} is mapped by $\mathfrak{T}$ to the lower right particle, which belongs to the other IR $\alpha = - 1 / 2$.
Similarly, since $W_{+ 1 / 2}^{\mathfrak{C}} = 0$ [Eq.~\eqref{eq:Wigner_criterion_C2_Ag_C}], the lower left particle is mapped by $\mathfrak{C}$ to the upper right hole.
On the other hand, since the orthogonality test leads to $W_{+ 1 / 2}^\Gamma = 1$ [Eq.~\eqref{eq:orthogonality_test_C2_Ag_G}], the CS gives the basis of the equivalent IR (see Appendix~\ref{sec:orthogonality_test}).
Thus the lower left particle in Fig.~\ref{fig:emergent_AZ_class_C2_Ag} is mapped by $\Gamma$ to the upper left hole, which belongs to the same IR $\alpha = + 1 / 2$.
For the above reason, the Hamiltonian in the space of $\bar{\gamma}^{\bm{k}}_{+ 1 / 2}$ (the red frame in Fig.~\ref{fig:emergent_AZ_class_C2_Ag}) has only the CS $\Gamma$, which indicates the AZ class AIII~\cite{Altland1997}.

Furthermore, we see an intuitive understanding of the gap opening in the $A_g$ symmetry.
Within the weak-coupling limit, i.e., for the negligibly small inter-band pairing, it is sufficient to discuss the single-band model.
In this case, the BdG Hamiltonian $H_{\text{BdG}}(\bm{k})$ on the $C_2$-symmetric line is generally written as follows:
\begin{align}
 H_{\text{BdG}}(\bm{k}) &= \frac{1}{2} \bm{C}^\dagger(\bm{k}) \hat{H}_{\text{BdG}}(\bm{k}) \bm{C}(\bm{k}), \label{eq:BdG_Hamiltonian} \\
 \bm{C}^\dagger(\bm{k}) &= (c_{\bm{k}, + 1 / 2}^\dagger, \, \mathfrak{T} c_{\bm{k}, + 1 / 2}^\dagger \mathfrak{T}^{-1}, \notag \\
 & \qquad \mathfrak{C} c_{\bm{k}, + 1 / 2}^\dagger \mathfrak{C}^{-1}, \, \Gamma c_{\bm{k}, + 1 / 2}^\dagger \Gamma^{-1}),
\end{align}
where $c_{\bm{k}, + 1 / 2}^\dagger$ is a creation operator of a $j_z = + 1 / 2$ Bloch state in a band-based representation, which is the basis of the IR $\bar{\gamma}^{\bm{k}}_{+ 1 / 2}$:
\begin{equation}
 g c_{\bm{k}, + 1 / 2}^\dagger g^{-1} = \bar{\gamma}^{\bm{k}}_{+ 1 / 2}(g) c_{\bm{k}, + 1 / 2}^\dagger \ \text{for} \ g \in \bar{\cal G}^{\bm{k}}.
\end{equation}
$\hat{H}_{\text{BdG}}(\bm{k})$ is the matrix representation of $H_{\text{BdG}}(\bm{k})$:
\begin{equation}
 \hat{H}_{\text{BdG}}(\bm{k}) =
  \begin{pmatrix}
   \xi_{\bm{k}} \hat{\sigma}_0 & \Delta_0 \psi_{\bm{k}} i\hat{\sigma}_2 \\
   (\Delta_0 \psi_{\bm{k}} i\hat{\sigma}_2)^\dagger & - \xi_{\bm{k}} \hat{\sigma}_0
  \end{pmatrix}, \label{eq:BdG_Hamiltonian_C2}
\end{equation}
where $\xi_{\bm{k}}$ is the normal energy dispersion, and $\hat{\sigma}_i$ ($i = 0, 1, 2, 3$) represents the Pauli matrix in pseudospin space.
Here, the \textit{spin-singlet} $A_g$ gap function is defined as
\begin{align}
 & \frac{1}{2} \Delta_0 \psi_{\bm{k}} \left\{ c_{\bm{k}, + 1 / 2}^\dagger (\Gamma c_{\bm{k}, + 1 / 2}^\dagger \Gamma^{-1})^\dagger \right. \notag \\
 & \qquad\qquad \left. - \mathfrak{T} c_{\bm{k}, + 1 / 2}^\dagger \mathfrak{T}^{-1} (\mathfrak{C} c_{\bm{k}, + 1 / 2}^\dagger \mathfrak{C}^{-1})^\dagger \right\} + \text{H.c.}, \label{eq:spin-singlet_gap}
\end{align}
where the magnitude of the gap function $\Delta_0$ is chosen as a real number without loss of generality.

Due to the $C_2$ symmetry, the BdG Hamiltonian matrix commutes with the twofold rotation matrix $\hat{U}_{\text{BdG}}^{C_2}$:
\begin{equation}
 [\hat{H}_{\text{BdG}}(\bm{k}), \hat{U}_{\text{BdG}}^{C_2}] = 0.
\end{equation}
Therefore $\hat{H}_{\text{BdG}}(\bm{k})$ and $\hat{U}_{\text{BdG}}^{C_2}$ are simultaneously block-diagonalized; namely, there exists a unitary matrix $\hat{V}$ such that
\begin{align}
 \hat{H}_{\text{BdG}}(\bm{k}) &= \hat{V} \begin{pmatrix}
                                          \hat{H}_+(\bm{k}) & 0 \\
                                          0 & \hat{H}_-(\bm{k})
                                         \end{pmatrix} \hat{V}^\dagger, \label{eq:block-diagonalized_Hamiltonian_C2} \\
 \hat{U}_{\text{BdG}}^{C_2} &= \hat{V} \begin{pmatrix}
                                        + i \bm{1}_2 & 0 \\
                                        0 & - i \bm{1}_2
                                       \end{pmatrix} \hat{V}^\dagger. \label{eq:block-diagonalized_C2}
\end{align}
The block-diagonalized Hamiltonian $H_+(\bm{k})$ and $H_-(\bm{k})$ are written as follows:
\begin{align}
 H_\pm(\bm{k}) &= \frac{1}{2} \bm{C}_\pm^\dagger(\bm{k}) \hat{H}_\pm(\bm{k}) \bm{C}_\pm(\bm{k}), \\
 \hat{H}_\pm(\bm{k}) &= \begin{pmatrix}
                         \xi_{\bm{k}} & \pm \Delta_0 \psi_{\bm{k}} \\
                         \pm \Delta_0 \psi_{\bm{k}}^* & - \xi_{\bm{k}}
                        \end{pmatrix}, \label{eq:BdG_Hamiltonian_C2_Ag} \\
 \bm{C}_+^\dagger(\bm{k}) &= (c_{\bm{k}, + 1 / 2}^\dagger, \, \Gamma c_{\bm{k}, + 1 / 2}^\dagger \Gamma^{-1}), \\
 \bm{C}_-^\dagger(\bm{k}) &= (\mathfrak{T} c_{\bm{k}, + 1 / 2}^\dagger \mathfrak{T}^{-1}, \, \mathfrak{C} c_{\bm{k}, + 1 / 2}^\dagger \mathfrak{C}^{-1}),
\end{align}
since $\Gamma$ does not change the eigenvalue of $\hat{U}_{\text{BdG}}^{C_2}$, but $\mathfrak{T}$ and $\mathfrak{C}$ do that.

\begin{figure*}[tbp]
 \centering
 \includegraphics[width=16cm, clip]{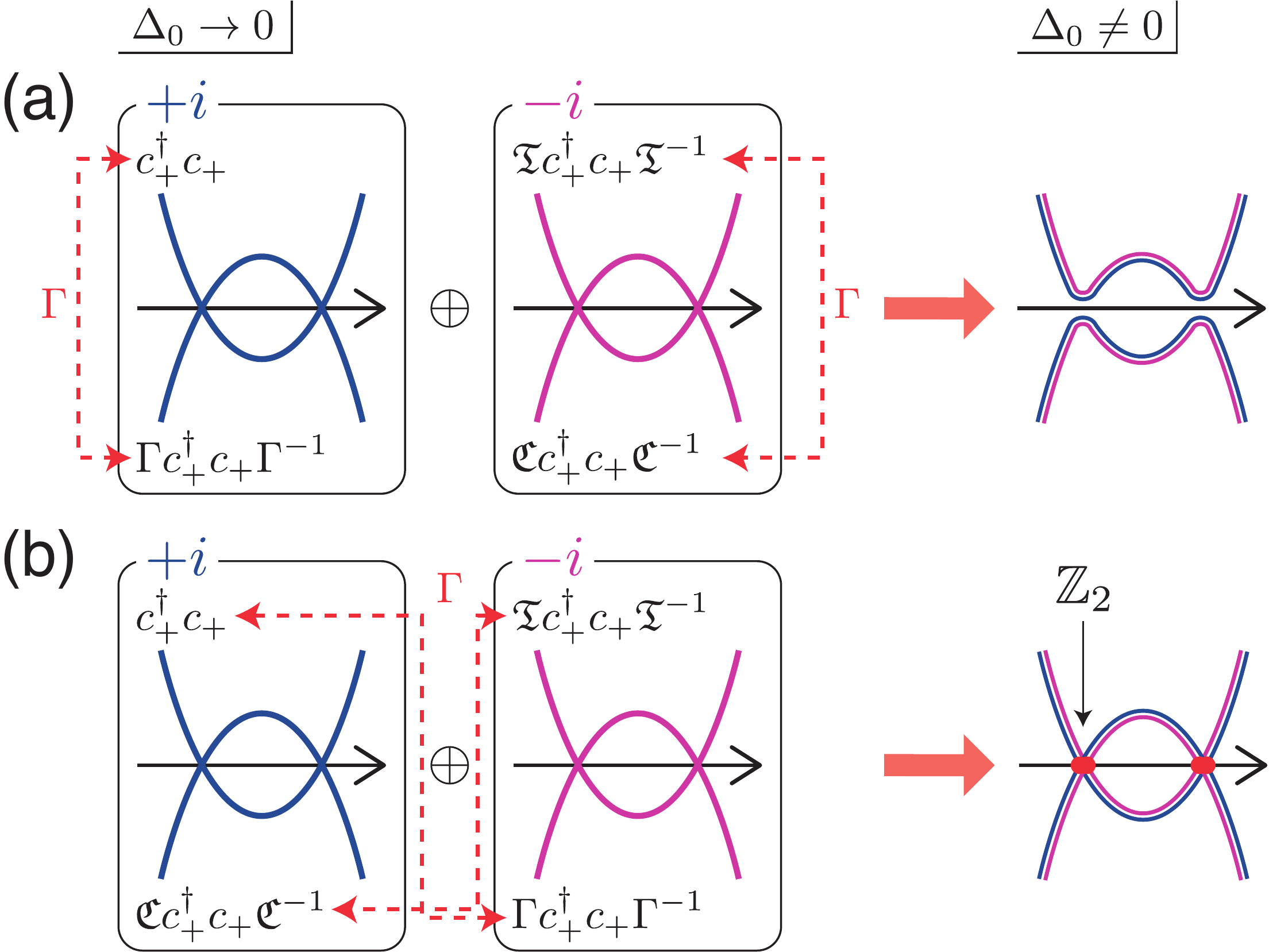}
 \caption{Band-theoretical picture of the BdG Hamiltonian with (a) the $A_g$ pair wave function and (b) the $B_g$ pair wave function on a $C_2$-symmetric line. $c_+^\dagger$ is the abbreviated notation of $c_{\bm{k}, + 1 / 2}^\dagger$. The red points represent nodes on the line.}
 \label{fig:eigenvalue_C2}
\end{figure*}

Figure~\ref{fig:eigenvalue_C2}(a) schematically illustrates the band structures obtained by the BdG Hamiltonian~\eqref{eq:block-diagonalized_Hamiltonian_C2}.
First, considering $\Delta_0 \to 0$ limit [left panel in Fig.~\ref{fig:eigenvalue_C2}(a)], we get the particle band
\begin{equation}
 \xi_{\bm{k}} c_{\bm{k}, + 1 / 2}^\dagger c_{\bm{k}, + 1 / 2} \quad (\xi_{\bm{k}} \mathfrak{T} c_{\bm{k}, + 1 / 2}^\dagger c_{\bm{k}, + 1 / 2} \mathfrak{T}^{-1}),
\end{equation}
and the hole band
\begin{equation}
 - \xi_{\bm{k}} \Gamma c_{\bm{k}, + 1 / 2}^\dagger c_{\bm{k}, + 1 / 2} \Gamma^{-1} \quad (- \xi_{\bm{k}} \mathfrak{C} c_{\bm{k}, + 1 / 2}^\dagger c_{\bm{k}, + 1 / 2} \mathfrak{C}^{-1}),
\end{equation}
in the eigenspace of the eigenvalue $+ i$ ($- i$).
When the magnitude of the gap function $\Delta_0$ is finite, therefore, the $A_g$ pair wave function [Eq.~\eqref{eq:spin-singlet_gap}] can have finite off-diagonal components in each $+ i$ and $- i$ eigenspace [see Eq.~\eqref{eq:BdG_Hamiltonian_C2_Ag}].
These functions open the gaps at the zero energy [right panel in Fig.~\ref{fig:eigenvalue_C2}(a)].
This fact indicates the existence of gap on the $C_2$-symmetric axis.

\subsubsection{Node on a \texorpdfstring{$C_2$}{C2} line}
Next, we investigate the $B_g$ pair wave function on the $C_2$-symmetric axis.
Group-theoretical gap classification shows the emergence of nodes on the axis (see Table~\ref{tab:gap_classification_group_theory}).

The little cogroup $\bar{\cal G}^{\bm{k}}$ and its IR matrices $\bar{\gamma}^{\bm{k}}_{\pm 1 / 2}$ have been given by Eqs.~\eqref{eq:little_cogroup_C2} and \eqref{eq:Bloch_rep_C2}, respectively.
We again choose the IR $\bar{\gamma}^{\bm{k}}_{+ 1 / 2}$ and apply the topological classification.
In this symmetry, the (anti-)commutation relations are given by
\begin{gather}
 \mathfrak{T}^2 = - E, \, [\mathfrak{T}, C_2] = 0, \label{eq:commutation_C2_Bg_T} \\
 \mathfrak{C}^2 = + E, \, \{\mathfrak{C}, C_2\} = 0, \label{eq:commutation_C2_Bg_C} \\
 \{\Gamma, C_2\} = - [\mathfrak{T}, C_2] \mathfrak{C} + \mathfrak{T} \{\mathfrak{C}, C_2\} = 0. \label{eq:commutation_C2_Bg_G}
\end{gather}
By using the above relationships, the Wigner criteria for $\mathfrak{T}$ and $\mathfrak{C}$ and the orthogonality test for $\Gamma$ are given by
\begin{align}
 W_{+ 1 / 2}^{\mathfrak{T}} &= \frac{1}{2} \left\{ z_{\mathfrak{T}, \mathfrak{T}}^{\bm{k}} \chi[\bar{\gamma}^{\bm{k}}_{+ 1 / 2}(\mathfrak{T}^2)] \right. \notag \\
 &\qquad \left. + z_{\mathfrak{T} C_2, \mathfrak{T} C_2}^{\bm{k}} \chi[\bar{\gamma}^{\bm{k}}_{+ 1 / 2}((\mathfrak{T} C_2)^2)] \right\} \notag \\
 &= \frac{1}{2} \{ - 1 + 1 \} = 0, \label{eq:Wigner_criterion_C2_Bg_T} \\
 W_{+ 1 / 2}^{\mathfrak{C}} &= \frac{1}{2} \left\{ z_{\mathfrak{C}, \mathfrak{C}}^{\bm{k}} \chi[\bar{\gamma}^{\bm{k}}_{+ 1 / 2}(\mathfrak{C}^2)] \right. \notag \\
 &\qquad \left. + z_{\mathfrak{C} C_2, \mathfrak{C} C_2}^{\bm{k}} \chi[\bar{\gamma}^{\bm{k}}_{+ 1 / 2}((\mathfrak{C} C_2)^2)] \right\} \notag \\
 &= \frac{1}{2} \{ + 1 + 1 \} = + 1, \label{eq:Wigner_criterion_C2_Bg_C} \\
 W_{+ 1 / 2}^\Gamma &= \frac{1}{2} \left\{ \frac{z_{E, \Gamma}^{\bm{k}}}{z_{\Gamma, \Gamma^{-1} E \Gamma}^{\bm{k}}} \chi[\bar{\gamma}^{\bm{k}}_{+ 1 / 2}(E)^*] \chi[\bar{\gamma}^{\bm{k}}_{+ 1 / 2}(E)] \right. \notag \\
 &\qquad \left. + \frac{z_{C_2, \Gamma}^{\bm{k}}}{z_{\Gamma, \Gamma^{-1} C_2 \Gamma}^{\bm{k}}} \chi[\bar{\gamma}^{\bm{k}}_{+ 1 / 2}(\Gamma^{-1} C_2 \Gamma)^*] \chi[\bar{\gamma}^{\bm{k}}_{+ 1 / 2}(C_2)] \right\} \notag \\
 &= \frac{1}{2} \{ 1 - 1 \} = 0. \label{eq:orthogonality_test_C2_Bg_G}
\end{align}
According to Table~\ref{tab:emergent_AZ_class}, therefore, the system is identified as the EAZ symmetry class D.
Since the class D is classified into $\mathbb{Z}_2$, nodes emerge on the $C_2$-symmetric line by the $B_g$ pair wave function when the FSs cross the line.
This node is topologically protected.

\begin{figure}[tbp]
 \centering
 \includegraphics[width=8cm, clip]{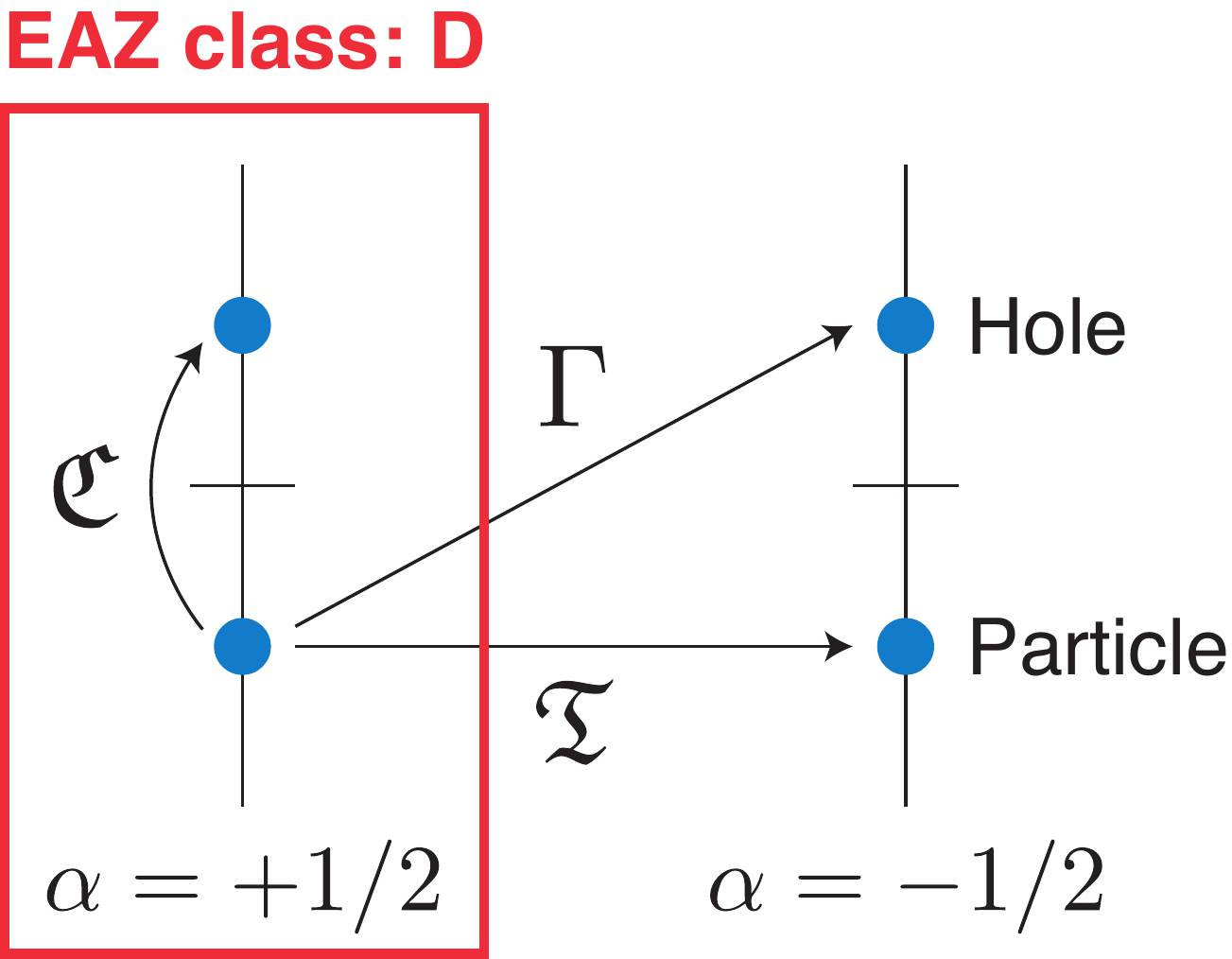}
 \caption{Schematic picture of the BdG Hamiltonian with the $B_g$ pair wave function on a $C_2$-symmetric line. The space in the red frame belongs to the EAZ class D.}
 \label{fig:emergent_AZ_class_C2_Bg}
\end{figure}

Figure~\ref{fig:emergent_AZ_class_C2_Bg} represents the schematic picture of the BdG Hamiltonian with the $B_g$ pair wave function.
The normal Bloch basis of $\bar{\gamma}^{\bm{k}}_{+ 1 / 2}$, namely, the lower left particle in Fig.~\ref{fig:emergent_AZ_class_C2_Bg}, is mapped by $\mathfrak{T}$ to the lower right particle belonging to the other IR $\alpha = - 1 / 2$, because the Wigner criterion for the TRS $\mathfrak{T}$ is $W_{+ 1 / 2}^{\mathfrak{T}} = 0$ [Eq.~\eqref{eq:Wigner_criterion_C2_Bg_T}].
Similarly, since $W_{+ 1 / 2}^{\Gamma} = 0$ [Eq.~\eqref{eq:orthogonality_test_C2_Bg_G}], the lower left particle is mapped by $\Gamma$ to the upper right hole.
On the other hand, since the Wigner criterion for the PHS is $W_{+ 1 / 2}^{\mathfrak{C}} = + 1$ [Eq.~\eqref{eq:Wigner_criterion_C2_Bg_C}], $\mathfrak{C}$ gives the basis of the equivalent IR which does \textit{not} generate an additional degeneracy (see Appendix~\ref{sec:Wigner_criterion}).
Thus the lower left particle in Fig.~\ref{fig:emergent_AZ_class_C2_Bg} is mapped by $\mathfrak{C}$ to the upper left hole belonging to the same IR $\alpha = + 1 / 2$.
For the above reason, the BdG Hamiltonian in the space of $\bar{\gamma}^{\bm{k}}_{+ 1 / 2}$ (the red frame in Fig.~\ref{fig:emergent_AZ_class_C2_Bg}) has only the PHS $\mathfrak{C}$ with $\mathfrak{C}^2 = + E$, which indicates the AZ class D~\cite{Altland1997}.

Furthermore, we discuss an intuitive picture of the nodes appearing in the $B_g$ symmetry.
As is the case for the $A_g$ IR, the BdG Hamiltonian matrix $\hat{H}_{\text{BdG}}(\bm{k})$ [Eq.~\eqref{eq:BdG_Hamiltonian_C2}] and the $C_2$ rotation matrix $\hat{U}_{\text{BdG}}^{C_2}$ are simultaneously block-diagonalized on the $C_2$-symmetric line [Eqs.~\eqref{eq:block-diagonalized_Hamiltonian_C2} and \eqref{eq:block-diagonalized_C2}].
For the $B_g$ IR, the Hamiltonian blocks $H_+(\bm{k})$ and $H_-(\bm{k})$ are written as follows:
\begin{align}
 H_\pm(\bm{k}) &= \frac{1}{2} \bm{C}_\pm^\dagger(\bm{k}) \hat{H}_\pm(\bm{k}) \bm{C}_\pm(\bm{k}), \\
 \hat{H}_\pm(\bm{k}) &= \begin{pmatrix}
                         \xi_{\bm{k}} & 0 \\
                         0 & - \xi_{\bm{k}}
                        \end{pmatrix}, \\
 \bm{C}_+^\dagger(\bm{k}) &= (c_{\bm{k}, + 1 / 2}^\dagger, \, \mathfrak{C} c_{\bm{k}, + 1 / 2}^\dagger \mathfrak{C}^{-1}), \\
 \bm{C}_-^\dagger(\bm{k}) &= (\mathfrak{T} c_{\bm{k}, + 1 / 2}^\dagger \mathfrak{T}^{-1}, \, \Gamma c_{\bm{k}, + 1 / 2}^\dagger \Gamma^{-1}),
\end{align}
since $\mathfrak{C}$ does not change the eigenvalue of $\hat{U}_{\text{BdG}}^{C_2}$, but $\mathfrak{T}$ and $\Gamma$ do that.
Note that the off-diagonal components of $\hat{H}_\pm(\bm{k})$ are zero because the \textit{spin-singlet} $B_g$ gap function has the same form as Eq.~\eqref{eq:spin-singlet_gap}, which is \textit{not} allowed in the equivalent eigenspace of $\pm i$.
In other words, the momentum dependence of the $B_g$ gap function $\psi_{\bm{k}}$ leads to the gap closing on the $C_2$-symmetric line, even when the magnitude $\Delta_0$ is finite.
The band structures are schematically shown in Fig.~\ref{fig:eigenvalue_C2}(b), which indicates the emergence of nodes on the $C_2$-symmetric axis [right panel in Fig.~\ref{fig:eigenvalue_C2}(b)].

Finally, we comment on the $\mathbb{Z}_2$ classification of the gap node.
As shown in Eq.~\eqref{eq:block-diagonalized_Hamiltonian_C2}, the BdG Hamiltonian matrix in this symmetry can be decomposed into the two matrices belonging to different eigenspaces of $\pm i$: $\hat{H}_+(\bm{k}) \oplus \hat{H}_-(\bm{k})$.
Both matrices possess the PHS $\mathfrak{C}$ with $\mathfrak{C}^2 = + E$.
Therefore $\hat{H}_\pm(\bm{k})$ can be transformed into antisymmetric matrices by using the unitary part $\hat{U}_{\mathfrak{C}} = \hat{U}_{\mathfrak{C}, +} \oplus \hat{U}_{\mathfrak{C}, -}$ of the PHS $\mathfrak{C}$~\cite{Ghosh2010, Agterberg2017}.
Thus the nodes in each eigenspace are characterized by a $\mathbb{Z}_2$ number:
\begin{equation}
 (-1)^{l_\pm} = \sgn[i^n \Pf\{\hat{U}_{\mathfrak{C}, \pm} \hat{H}_\pm(\bm{k})\}] \in \mathbb{Z}_2,
\end{equation}
with $n = \dim(\hat{H}_\pm) / 2$.
This $\mathbb{Z}_2$ protection of nodes is the same as that of Bogoliubov FSs in even-parity chiral superconductors~\cite{Agterberg2017}.

\subsection{Complete classification on \texorpdfstring{$C_n$ or $C_{nv}$}{Cn or Cnv} lines}
As revealed by the above two examples, the gap (node) on a high-symmetry line is represented by the absence (presence) of a topological number in the classification by the EAZ symmetry class.
Similarly, we can classify all superconducting gap structures on $C_n$- or $C_{nv}$-symmetric lines in the BZ, which are summarized in Table~\ref{tab:gap_classification_topological}.
Note that the EAZ classes in the table are \textit{not} equal to the AZ symmetry classes of the \textit{total} BdG Hamiltonian, but represent the symmetry of the Hamiltonian decomposed by the IRs of $\bar{\cal G}^{\bm{k}}$.
From Table~\ref{tab:gap_classification_topological}, we can identify whether the superconducting gap closes or not on \textit{almost all the high-symmetric lines} with some exceptions\footnote{For example, the gap classification on a $C_{2v}$-symmetric hinge of the BZ with \textit{glide symmetry} is different from that of Table~\ref{tab:gap_classification_topological}(e) with mirror symmetry~\cite{Yoshida2018_arXiv}.} by determining the IR of the normal Bloch state and that of the superconducting order parameter.

\begin{table*}[tbp]
 \caption{Topological classification of gap structures on high-symmetry lines in the BZ. Each classification is represented by the type of the topological number and the gap structure on the line: (G) full gap, (P) point nodes, (L) a part of line nodes, and (S) a part of surface nodes (Bogoliubov FSs). In a spontaneously TRS breaking phase, since all 2D IRs are decomposed into the 1D IRs with different eigenvalues of the rotation symmetry (see also Table~\ref{tab:character_2D_IRs}), the IRs in $D_{3d}$, $D_{4h}$, and $D_{6h}$ are the same as those in $S_6$, $C_{4h}$, and $C_{6h}$, respectively. Therefore we do not show the 2D IRs in the tables (f), (g), and (h).}
 \label{tab:gap_classification_topological}
 \begin{center}
  \begin{tabular}{cccp{10mm}cccp{5mm}ccc} \hline\hline
   \multicolumn{3}{c}{(a) $\bar{\cal G}^{\bm{k}} = C_2$, $\alpha = \pm 1 / 2$} & & \multicolumn{3}{c}{(b1) $\bar{\cal G}^{\bm{k}} = C_3$, $\alpha = + [-] 1 / 2$} & \quad & \multicolumn{3}{c}{(b2) $\bar{\cal G}^{\bm{k}} = C_3$, $\alpha = \pm 3 / 2$} \\ \cline{1-3}\cline{5-7}\cline{9-11}
   IR of $C_{2h}$ & EAZ & Classification & & IR of $S_6$ & EAZ & Classification & & IR of $S_6$ & EAZ & Classification \\ \cline{1-3}\cline{5-11}
   $A_g$ & AIII & $0$ (G)            & & $A_g$               & AIII & $0$ (G)            & & $A_g$        & DIII & $0$ (G)          \\
   $A_u$ & AIII & $0$ (G)            & & $A_u$               & AIII & $0$ (G)            & & $A_u$        & CII  & $0$ (G)          \\
   $B_g$ & D    & $\mathbb{Z}_2$ (L) & & $^{2}E_g [^{1}E_g]$ & D    & $\mathbb{Z}_2$ (S) & & $^{1, 2}E_g$ & A    & $\mathbb{Z}$ (S) \\
   $B_u$ & C    & $0$ (G)            & & $^{1}E_g [^{2}E_g]$ & A    & $\mathbb{Z}$ (S)   & &              &      &                  \\
         &      &                    & & $^{2}E_u [^{1}E_u]$ & C    & $0$ (G)            & & $^{1, 2}E_u$ & A    & $\mathbb{Z}$ (P) \\
         &      &                    & & $^{1}E_u [^{2}E_u]$ & A    & $\mathbb{Z}$ (P)   & &              &      &                  \\
   \\
   \multicolumn{3}{c}{(c) $\bar{\cal G}^{\bm{k}} = C_4$, $\alpha = + [-] 1 / 2, + [-] 3 / 2$} & & \multicolumn{3}{c}{(d1) $\bar{\cal G}^{\bm{k}} = C_6$, $\alpha = + [-] 1 / 2, + [-] 5 / 2$} & & \multicolumn{3}{c}{(d2) $\bar{\cal G}^{\bm{k}} = C_6$, $\alpha = \pm 3 / 2$} \\ \cline{1-3}\cline{5-7}\cline{9-11}
  IR of $C_{4h}$ & EAZ & Classification & & IR of $C_{6h}$ & EAZ & Classification & & IR of $C_{6h}$ & EAZ & Classification \\ \cline{1-3}\cline{5-11}
   $A_g$               & AIII & $0$ (G)            & & $A_g$                     & AIII & $0$ (G)            & & $A_g$           & AIII & $0$ (G)            \\
   $A_u$               & AIII & $0$ (G)            & & $A_u$                     & AIII & $0$ (G)            & & $A_u$           & AIII & $0$ (G)            \\
   $B_g$               & A    & $\mathbb{Z}$ (L)   & & $B_g$                     & A    & $\mathbb{Z}$ (L)   & & $B_g$           & D    & $\mathbb{Z}_2$ (L) \\
   $B_u$               & A    & $\mathbb{Z}$ (P)   & & $B_u$                     & A    & $\mathbb{Z}$ (P)   & & $B_u$           & C    & $0$ (G)            \\
   $^{2}E_g [^{1}E_g]$ & D    & $\mathbb{Z}_2$ (S) & & $^{1}E_{1g} [^{2}E_{1g}]$ & D    & $\mathbb{Z}_2$ (S) & & $^{1, 2}E_{1g}$ & A    & $\mathbb{Z}$ (S)   \\
   $^{1}E_g [^{2}E_g]$ & A    & $\mathbb{Z}$ (S)   & & $^{2}E_{1g} [^{1}E_{1g}]$ & A    & $\mathbb{Z}$ (S)   & &                 &      &                    \\
   $^{2}E_u [^{1}E_u]$ & C    & $0$ (G)            & & $^{1}E_{1u} [^{2}E_{1u}]$ & C    & $0$ (G)            & & $^{1, 2}E_{1u}$ & A    & $\mathbb{Z}$ (P)   \\
   $^{1}E_u [^{2}E_u]$ & A    & $\mathbb{Z}$ (P)   & & $^{2}E_{1u} [^{1}E_{1u}]$ & A    & $\mathbb{Z}$ (P)   & &                 &      &                    \\
                       &      &                    & & $^{1, 2}E_{2g}$           & A    & $\mathbb{Z}$ (S)   & & $^{1, 2}E_{2g}$ & A    & $\mathbb{Z}$ (S)   \\
                       &      &                    & & $^{1, 2}E_{2u}$           & A    & $\mathbb{Z}$ (P)   & & $^{1, 2}E_{2u}$ & A    & $\mathbb{Z}$ (P)   \\
   \\
   \multicolumn{3}{c}{(e) $\bar{\cal G}^{\bm{k}} = C_{2v}$, $\alpha = 1 / 2$} & & \multicolumn{3}{c}{(f1) $\bar{\cal G}^{\bm{k}} = C_{3v}$, $\alpha =  1 / 2$} & \quad & \multicolumn{3}{c}{(f2) $\bar{\cal G}^{\bm{k}} = C_{3v}$, $\alpha = 3 / 2$} \\ \cline{1-3}\cline{5-7}\cline{9-11}
   IR of $D_{2h}$ & EAZ & Classification & & IR of $D_{3d}$ & EAZ & Classification & & IR of $D_{3d}$ & EAZ & Classification \\ \cline{1-3}\cline{5-11}
   $A_g$    & CI  & $0$ (G)            & & $A_{1g}$ & CI       & $0$ (G)            & & $A_{1g}$ & AIII     & $0$ (G)            \\
   $A_u$    & CI  & $0$ (G)            & & $A_{1u}$ & CI       & $0$ (G)            & & $A_{1u}$ & C        & $0$ (G)            \\
   $B_{1g}$ & BDI & $\mathbb{Z}_2$ (L) & & $A_{2g}$ & BDI      & $\mathbb{Z}_2$ (L) & & $A_{2g}$ & D        & $\mathbb{Z}_2$ (L) \\
   $B_{1u}$ & BDI & $\mathbb{Z}_2$ (P) & & $A_{2u}$ & BDI      & $\mathbb{Z}_2$ (P) & & $A_{2u}$ & AIII     & $0$ (G)            \\
   $B_{2g}$ & BDI & $\mathbb{Z}_2$ (L) & & 2D IRs   & see (b1) &                    & & 2D IRs   & see (b2) &                    \\
   $B_{2u}$ & CI  & $0$ (G)            & &                                                                                       \\
   $B_{3g}$ & BDI & $\mathbb{Z}_2$ (L) & &                                                                                       \\
   $B_{3u}$ & CI  & $0$ (G)            & &                                                                                       \\
   \\
   \multicolumn{3}{c}{(g) $\bar{\cal G}^{\bm{k}} = C_{4v}$, $\alpha = 1 / 2, 3 / 2$} & & \multicolumn{3}{c}{(h1) $\bar{\cal G}^{\bm{k}} = C_{6v}$, $\alpha = 1 / 2, 5 / 2$} & & \multicolumn{3}{c}{(h2) $\bar{\cal G}^{\bm{k}} = C_{6v}$, $\alpha = 3 / 2$} \\ \cline{1-3}\cline{5-7}\cline{9-11}
   IR of $D_{4h}$ & EAZ & Classification & & IR of $D_{6h}$ & EAZ & Classification & & IR of $D_{6h}$ & EAZ & Classification \\ \cline{1-3}\cline{5-11}
   $A_{1g}$ & CI      & $0$ (G)            & & $A_{1g}$ & CI       & $0$ (G)            & & $A_{1g}$ & CI       & $0$ (G)            \\
   $A_{1u}$ & CI      & $0$ (G)            & & $A_{1u}$ & CI       & $0$ (G)            & & $A_{1u}$ & CI       & $0$ (G)            \\
   $A_{2g}$ & BDI     & $\mathbb{Z}_2$ (L) & & $A_{2g}$ & BDI      & $\mathbb{Z}_2$ (L) & & $A_{2g}$ & BDI      & $\mathbb{Z}_2$ (L) \\
   $A_{2u}$ & BDI     & $\mathbb{Z}_2$ (P) & & $A_{2u}$ & BDI      & $\mathbb{Z}_2$ (P) & & $A_{2u}$ & BDI      & $\mathbb{Z}_2$ (P) \\
   $B_{1g}$ & AI      & $\mathbb{Z}$ (L)   & & $B_{1g}$ & AI       & $\mathbb{Z}$ (L)   & & $B_{1g}$ & BDI      & $\mathbb{Z}_2$ (L) \\
   $B_{1u}$ & AI      & $\mathbb{Z}$ (P)   & & $B_{1u}$ & AI       & $\mathbb{Z}$ (P)   & & $B_{1u}$ & CI       & $0$ (G)            \\
   $B_{2g}$ & AI      & $\mathbb{Z}$ (L)   & & $B_{2g}$ & AI       & $\mathbb{Z}$ (L)   & & $B_{2g}$ & BDI      & $\mathbb{Z}_2$ (L) \\
   $B_{2u}$ & AI      & $\mathbb{Z}$ (P)   & & $B_{2u}$ & AI       & $\mathbb{Z}$ (P)   & & $B_{2u}$ & CI       & $0$ (G)            \\
   2D IRs   & see (c) &                    & & 2D IRs   & see (d1) &                    & & 2D IRs   & see (d2) &                    \\ \hline\hline
  \end{tabular}
 \end{center}
\end{table*}

Now we remark the treatment of the 2D IRs in Table~\ref{tab:gap_classification_topological}.
A general 2D superconducting order parameter matrix has the form $\hat{\Delta}(\bm{k}) = \eta_+ \hat{\Delta}_+(\bm{k}) + \eta_- \hat{\Delta}_-(\bm{k})$; for example, we consider a superconducting order parameter belonging to the $E_g$ IR of $C_3$, for which
\begin{equation}
 C_3 \hat{\Delta}_\pm(\bm{k}) C_3^T = e^{\pm i 2\pi / 3} \hat{\Delta}_\pm(\bm{k}), \label{eq:C3_order_pm}
\end{equation}
on the $C_3$-symmetric line.
For arbitrary parameters $\eta_+$ and $\eta_-$, however, it is impossible to determine a commutation relation between the rotation symmetry $C_3$ and the PHS ${\cal C}$.
On the other hand, if we choose the 1D order parameter with one of the two rotation-invariant bases $\hat{\Delta}(\bm{k}) = \hat{\Delta}_\pm(\bm{k})$, namely $(\eta_+, \eta_-) = (1, 0)$ or $(0, 1)$, the commutation can be given by
\begin{equation}
 C_3 {\cal C} = e^{\pm i 2\pi / 3} {\cal C} C_3. \label{eq:commutation_C3_C_Eg}
\end{equation}
In other words, $\hat{\Delta}(\bm{k}) = \hat{\Delta}_+(\bm{k})$ ($\hat{\Delta}_-(\bm{k})$) belongs to the 1D IR $^{2}E_g$ ($^{1}E_g$) of $S_6$ [see Table~\ref{tab:character_2D_IRs}(a)].

Note that since $\hat{\Delta}_+(\bm{k})$ and $\hat{\Delta}_-(\bm{k})$ are a pair of the order parameter connected by TRS, choosing one of them leads to TRS and CS breaking.
Thus we only have to calculate the Wigner criterion for the PHS-like operator $\mathfrak{C} = {\cal C I}$:
\begin{align}
 W_{+ 1 / 2}^{\mathfrak{C}} &= \frac{1}{3} \sum_{g \in \{E, C_3, (C_3)^2\}} z_{\mathfrak{C} g, \mathfrak{C} g}^{\bm{k}} \chi[\bar{\gamma}^{\bm{k}}_{+ 1 / 2}((\mathfrak{C} g)^2)] \notag \\
 &= \begin{cases}
     0 & ^{1}E_g, \\
     1 & ^{2}E_g,
    \end{cases} \label{eq:Wigner_criterion_C3_1/2_Eg_C} \\[5mm]
 W_{+ 3 / 2}^{\mathfrak{C}} &= \frac{1}{3} \sum_{g \in \{E, C_3, (C_3)^2\}} z_{\mathfrak{C} g, \mathfrak{C} g}^{\bm{k}} \chi[\bar{\gamma}^{\bm{k}}_{+ 3 / 2}((\mathfrak{C} g)^2)] \notag \\
 &= 0 \quad (^{1, 2}E_g), \label{eq:Wigner_criterion_C3_3/2_Eg_C}
\end{align}
where we use Eq.~\eqref{eq:commutation_C3_C_Eg}, $\mathfrak{C}^2 = + E$, $(C_3)^3 = - E$, and the IRs of $C_3$ in the following.
\begin{equation}
 \begin{array}{cccc} \hline\hline
  \text{IR} & E & C_3 & (C_3)^2 \\ \hline
  \bar{\gamma}^{\bm{k}}_{+ 1 / 2} & 1 & e^{+ i\pi / 3} & e^{+ i2\pi / 3} \\
  \bar{\gamma}^{\bm{k}}_{+ 3 / 2} & 1 & -1 & 1 \\ \hline\hline
 \end{array}
\end{equation}
Here the 1D IRs $^{1, 2}E_g$ of $S_6$ are defined in Table~\ref{tab:character_2D_IRs}(a).
Equations~\eqref{eq:Wigner_criterion_C3_1/2_Eg_C} and \eqref{eq:Wigner_criterion_C3_3/2_Eg_C} show that the classification results of $^{1}E_g$ and $^{2}E_g$ are different (equivalent) for the IR of the normal Bloch state $\alpha = + 1 / 2$ ($+ 3 / 2$); see also Tables~\ref{tab:gap_classification_topological}(b1) and \ref{tab:gap_classification_topological}(b2).
We remark that the calculations of Eq.~\eqref{eq:Wigner_criterion_C3_1/2_Eg_C} are reversed for an $\alpha = - 1 / 2$ band:
\begin{equation}
 W_{- 1 / 2}^{\mathfrak{C}} = 
  \begin{cases}
   1 & ^{1}E_g, \\
   0 & ^{2}E_g.
  \end{cases}
\end{equation}
These results are represented in the square brackets in Table~\ref{tab:gap_classification_topological}(b1).
In case of the other 2D IRs, we similarly decompose them into the 1D IRs with different eigenvalues of the rotation symmetry (see Table~\ref{tab:character_2D_IRs}).
For such 1D IRs produced from 2D IRs, therefore, the groups $D_{3d}$, $D_{4h}$, and $D_{6h}$ are reduced to $S_6$, $C_{4h}$, and $C_{6h}$, respectively.
Thus we do not show the 2D IRs for $D_{3d}$, $D_{4h}$, and $D_{6h}$ groups in Table~\ref{tab:gap_classification_topological}.

\begin{table}[tbp]
 \caption{Character tables for 2D IRs of the point groups (a) $S_6$, (b) $C_{4h}$, and (c) $C_{6h}$. All notations are based on Bilbao Crystallographic Server~\cite{Bilbao}. It is noteworthy that all representations labeled by the indices 1 and 2 are 1D IRs, which are doubly degenerated under TRS (e.g., $E_u = {}^{1}E_u + {}^{2}E_u$). In each table, characters are shown only for generators of the corresponding group.}
 \label{tab:character_2D_IRs}
 \begin{center}
  \begin{tabular}{ccccp{10mm}cccc} \hline\hline
   (a) $S_6$                & $E$ & $C_3$              & $I$   & & (c) $C_{6h}$                & $E$ & $C_6$               & $I$   \\ \hline
   $^{1}E_g$ \ $\Gamma_3^+$ & $1$ & $e^{- i 2\pi / 3}$ & $+ 1$ & & $^{1}E_{1g}$ \ $\Gamma_6^+$ & $1$ & $e^{- i \pi / 3}$   & $+ 1$ \\
   $^{2}E_g$ \ $\Gamma_2^+$ & $1$ & $e^{+ i 2\pi / 3}$ & $+ 1$ & & $^{2}E_{1g}$ \ $\Gamma_5^+$ & $1$ & $e^{+ i \pi / 3}$   & $+ 1$ \\
   $^{1}E_u$ \ $\Gamma_2^-$ & $1$ & $e^{- i 2\pi / 3}$ & $- 1$ & & $^{1}E_{1u}$ \ $\Gamma_6^-$ & $1$ & $e^{- i \pi / 3}$   & $- 1$ \\
   $^{2}E_u$ \ $\Gamma_3^-$ & $1$ & $e^{+ i 2\pi / 3}$ & $- 1$ & & $^{2}E_{1u}$ \ $\Gamma_5^-$ & $1$ & $e^{+ i \pi / 3}$   & $- 1$ \\
                            &     &                    &       & & $^{1}E_{2g}$ \ $\Gamma_3^+$ & $1$ & $- e^{- i \pi / 3}$ & $+ 1$ \\
   (b) $C_{4h}$             & $E$ & $C_4$              & $I$   & & $^{2}E_{2g}$ \ $\Gamma_2^+$ & $1$ & $- e^{+ i \pi / 3}$ & $+ 1$ \\ \cline{1-4}
   $^{1}E_g$ \ $\Gamma_4^+$ & $1$ & $- i$              & $+ 1$ & & $^{1}E_{2u}$ \ $\Gamma_3^-$ & $1$ & $- e^{- i \pi / 3}$ & $- 1$ \\
   $^{2}E_g$ \ $\Gamma_3^+$ & $1$ & $+ i$              & $+ 1$ & & $^{2}E_{2u}$ \ $\Gamma_2^-$ & $1$ & $- e^{+ i \pi / 3}$ & $- 1$ \\
   $^{1}E_u$ \ $\Gamma_4^-$ & $1$ & $- i$              & $- 1$ &                                                                  \\
   $^{2}E_u$ \ $\Gamma_3^-$ & $1$ & $+ i$              & $- 1$ &                                                                  \\ \hline\hline 
  \end{tabular}
 \end{center}
\end{table}

\begin{figure}[tbp]
 \centering
 \includegraphics[width=8cm, clip]{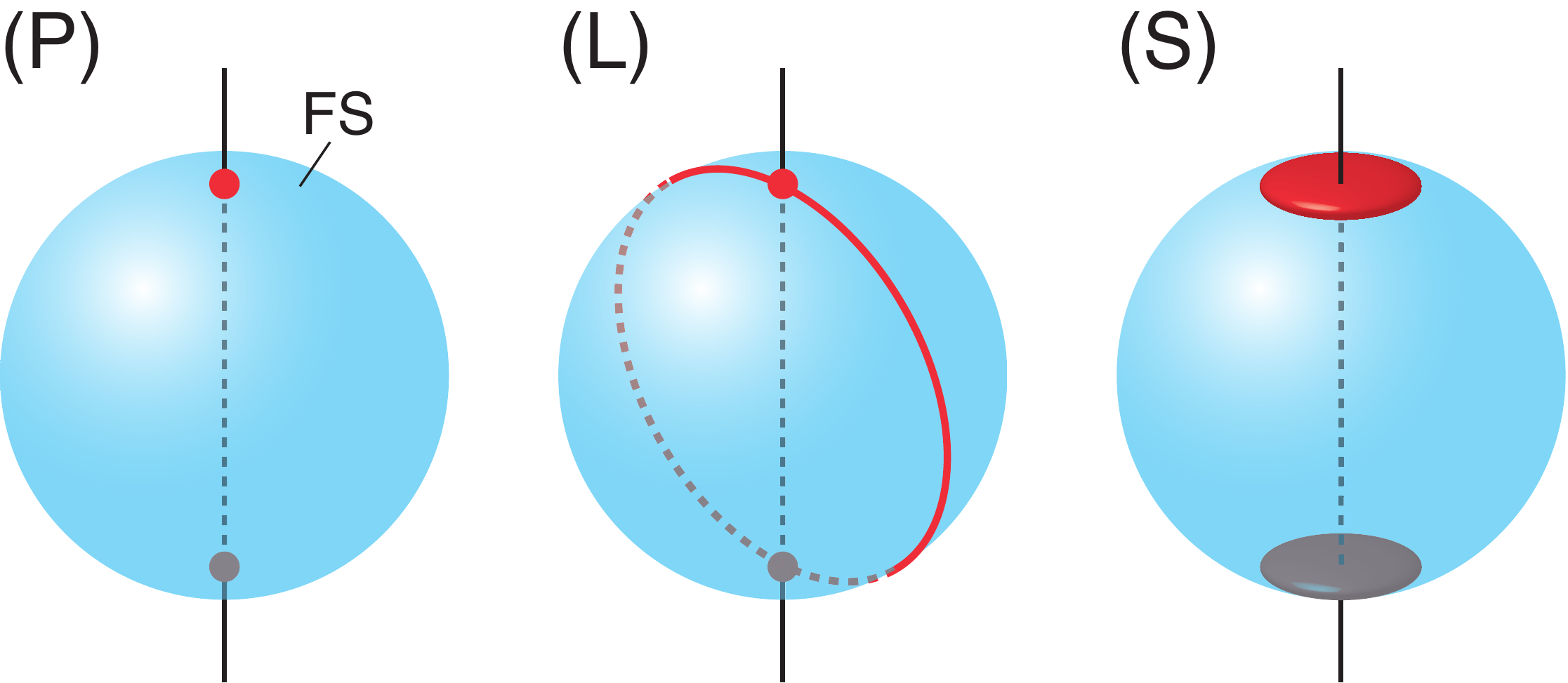}
 \caption{Three cases of the gap closing on a high-symmetry axis: (P) point nodes, (L) a part of line nodes, and (S) a part of surface nodes (Bogoliubov FSs).}
 \label{fig:node_type}
\end{figure}

In Table~\ref{tab:gap_classification_topological}, the classification ``$0$'' indicates the fully gapped structure on the intriguing high-symmetry line, which is consistent with the results of Table~\ref{tab:gap_classification_group_theory}.
If the classification is nontrivial ($\mathbb{Z}$ or $\mathbb{Z}_2$), on the other hand, the gap closes on the line.
As shown in Fig.~\ref{fig:node_type}, such gap structures have three types when the whole FS is considered: (P) point nodes, (L) a part of line nodes, and (S) a part of surface nodes (Bogoliubov FSs).
The condition for each node structure is specified by the parity and TRS of the order parameter as follows.

\textit{Case (S): even-parity and TRS breaking order parameter.}
When both particle bands and hole bands are doubly degenerate, nodes on the high-symmetry axis are point nodes.
Indeed, most of the single-band models reproduce the situation.
In real superconductors where multi-band effects cannot be neglected, however, the degeneracy splits due to the inter-band pairing effect~\cite{Agterberg2017}.
As a result, the point nodes are inflated to surface nodes, which are characterized by the 0D topological number (Pfaffian of the antisymmetrized total BdG Hamiltonian) $P(\bm{k}) \in \mathbb{Z}_2$~\cite{Agterberg2017}.

\textit{Case (L): even-parity and TRS preserving order parameter.}
When the high-symmetry axis possesses the mirror symmetry parallel to the axis (i.e., $\bar{\cal G}^{\bm{k}} = C_{nv}$), we can classify the presence or absence of line nodes on the mirror-invariant plane~\cite{Micklitz2017_PRL, Sumita2018, Kobayashi2018}.
Taking into account the compatibility relation between the high-symmetry axis and the mirror-invariant plane, it is found that nodes on the $C_{nv}$-symmetric axis are a part of line nodes on the plane.
The line nodes are also protected by the 1D topological number (winding number) $W_l \in \mathbb{Z}$, which is defined by the CS~\cite{Sato2011, Kobayashi2014, Kobayashi2018}.
Even when the mirror symmetry is broken, therefore, the line nodes do not vanish as long as the CS is preserved.
Since the line nodes remain intersectant with the axis because of the rotational symmetry, nodes on the $C_n$-symmetric axis are also a part of the line nodes.

\textit{Case (P): odd-parity order parameter.}
Except for the BZ boundary in nonsymmorphic space groups,\footnote{This case is beyond the scope of the paper.} there is no line node in odd-parity superconductivity~\cite{Blount1985, Kobayashi2014}.
\begin{enumerate}
 \item[(P1)] When the order parameter breaks TRS, band degeneracy generally splits.
             However, nodes on the high-symmetry axis are not a part of surface nodes since no 0D topological number is defined at a general point in the 3D BZ.
             Thus the stable nodes are point nodes.
 \item[(P2)] When the order parameter preserves TRS, nodes on the high-symmetry axis are point nodes because the band splitting does not occur.
\end{enumerate}

The classification of the types of nodes is shown in Table~\ref{tab:gap_classification_topological}.
Although a part of above discussions is speculative, we confirmed the validity of the results using the differential of Atiyah-Hirzebruch spectral sequence~\cite{Shiozaki2018_arXiv} except for physically unnatural cases.
The detailed results will be shown in our next publication~\cite{Nomoto2018_unpublished}.

\section{Candidate superconductors}
\label{sec:candidates}
In this section, we see some representative examples of symmetry-protected nodes on a high-symmetry line.
We introduce surface nodes in SrPtAs, line nodes in CeCoIn$_5$, and point nodes in UPt$_3$ and UCoGe.

\subsection{Surface nodes: SrPtAs with \texorpdfstring{$E_{2g}$}{E2g} order parameter}
\label{sec:SrPtAs}
\subsubsection{Background}
SrPtAs is a pnictide superconductor with a hexagonal lattice characterized by the nonsymmorphic space group $P6_3/mmc$ ($D_{6h}^4$).
First-principles studies using local density approximation show 2D FSs enclosing the $\Gamma$-$A$ line, a 2D FS enclosing the $K$-$H$ line, and a three-dimensional FS crossing the $K$-$H$ line~\cite{Shein2011, Youn2012, Youn2012_arXiv}.
The pairing symmetry of SrPtAs is still under debate because of the incompatibility of some experiments: for example, TRS breaking and a nodeless pairing gap suggested in a muon spin-rotation/relaxation measurement~\cite{Biswas2013} are incompatible with a spin-singlet $s$-wave superconducting state with an isotropic gap indicated by recent $^{195}$Pt-nuclear magnetic resonance and $^{75}$As-nuclear quadrupole resonance measurements~\cite{Matano2014}.

From the theoretical point of view, on the other hand, the $E_{2g}$ state with a chiral $d$-wave pairing~\cite{Goryo2012, Fischer2014, Fischer2015} and the $B_{1u}$ state with an $f$-wave pairing~\cite{WangW2014, Bruckner2014} have been proposed.
In the following discussion, we assume the chiral $E_{2g}$ state~\cite{Goryo2012, Fischer2014, Fischer2015}, which is consistent with broken TRS~\cite{Biswas2013} and the decrease of spin susceptibility below $T_{\text{c}}$~\cite{Matano2014}, but is incompatible with the nodeless gap structure~\cite{Biswas2013, Matano2014}.
Especially, we focus on the gap structure on the $C_{3v}$-symmetric $K$-$H$ line, since the presence of Weyl nodes~\cite{Fischer2014} or Bogoliubov FSs~\cite{Agterberg2017, Bzdusek2017, Brydon2018} have been suggested on the line.

\subsubsection{Classification on \texorpdfstring{$K$-$H$}{K-H} line}
Now we classify the gap structure on the $K$-$H$ line with the $C_{3v}$ symmetry.
The compatibility relation reveals that $E_{2g}$ of $D_{6h}$ corresponds to $E_g$ of $D_{3d}$.
According to Table~\ref{tab:gap_classification_group_theory}, $E_g$ gap closes on the line irrespective of the angular momentum of normal Bloch states.
However, Tables~\ref{tab:gap_classification_topological}(b1) and \ref{tab:gap_classification_topological}(b2) shows the distinct topological classifications of such nodes depending on the angular momentum: $\mathbb{Z}_2 \oplus \mathbb{Z}$ for $\alpha = \pm 1 / 2$ and $\mathbb{Z} \oplus \mathbb{Z}$ for $\alpha = \pm 3 / 2$.

Next, we concretely identify the topological number.
In the discussions below, we fix the superconducting order parameter to the $^{2}E_g$ IR, which obtains the phase factor $+ 2\pi / 3$ under a $C_3$ rotation:
\begin{equation}
 C_3 \hat{\Delta}_+(\bm{k}) C_3^T = e^{+ i 2\pi / 3} \hat{\Delta}_+(\bm{k}),
\end{equation}
for $\bm{k}$ on the $C_3$-symmetric line.
Some theoretical studies~\cite{Goryo2012, Fischer2014, Fischer2015} have suggested such a TRS breaking chiral order parameter.
Although only the PHS $\mathfrak{C}$ is preserved in this TRS (CS) breaking superconducting state, the TRS $\mathfrak{T}$ and the CS $\Gamma$ recover by considering $\Delta_0 \to 0$ limit.
Thus, in this limit, we can also define the BdG Hamiltonian on the $C_3$-symmetric line as Eq.~\eqref{eq:BdG_Hamiltonian} and
\begin{align}
 \bm{C}^\dagger(\bm{k}) &= (c_{\bm{k}, + 1 / 2}^\dagger, \, \mathfrak{T} c_{\bm{k}, + 1 / 2}^\dagger \mathfrak{T}^{-1}, & \notag \\
 & \quad \mathfrak{C} c_{\bm{k}, + 1 / 2}^\dagger \mathfrak{C}^{-1}, \, \Gamma c_{\bm{k}, + 1 / 2}^\dagger \Gamma^{-1}) & \text{for} \ \alpha = \pm 1 / 2, \\
 \bm{C}^\dagger(\bm{k}) &= (c_{\bm{k}, + 3 / 2}^\dagger, \, \mathfrak{T} c_{\bm{k}, + 3 / 2}^\dagger \mathfrak{T}^{-1}, & \notag \\
 & \quad \mathfrak{C} c_{\bm{k}, + 3 / 2}^\dagger \mathfrak{C}^{-1}, \, \Gamma c_{\bm{k}, + 3 / 2}^\dagger \Gamma^{-1}) & \text{for} \ \alpha = \pm 3 / 2.
\end{align}
The BdG Hamiltonian matrix is written by the following effective single-band model:
\begin{equation}
 \hat{H}_{\text{BdG}}(\bm{k}) =
  \begin{pmatrix}
   \xi_{\bm{k}} \hat{\sigma}_0 - h_{\bm{k}} \Delta_0^2 \hat{\sigma}_3 & \Delta_0 \psi_{\bm{k}} i\hat{\sigma}_2 \\
   (\Delta_0 \psi_{\bm{k}} i\hat{\sigma}_2)^\dagger & - (\xi_{\bm{k}} \hat{\sigma}_0 - h_{\bm{k}} \Delta_0^2 \hat{\sigma}_3)
  \end{pmatrix},
\end{equation}
where $h_{\bm{k}} \Delta_0^2$ is the ``pseudomagnetic'' field representing the TRS breaking in the superconducting ordered state.
The low-energy effective theory elucidates that such a field arises from the second-order perturbation of the \textit{inter-band} pairing~\cite{Venderbos2018, Brydon2018}, which cannot be generally neglected in real multiband systems.
Note that the pseudomagnetic field does not break $C_3$ symmetry since it is parallel to the $C_3$-symmetric axis.

\begin{figure*}[tbp]
 \centering
 \includegraphics[width=16cm, clip]{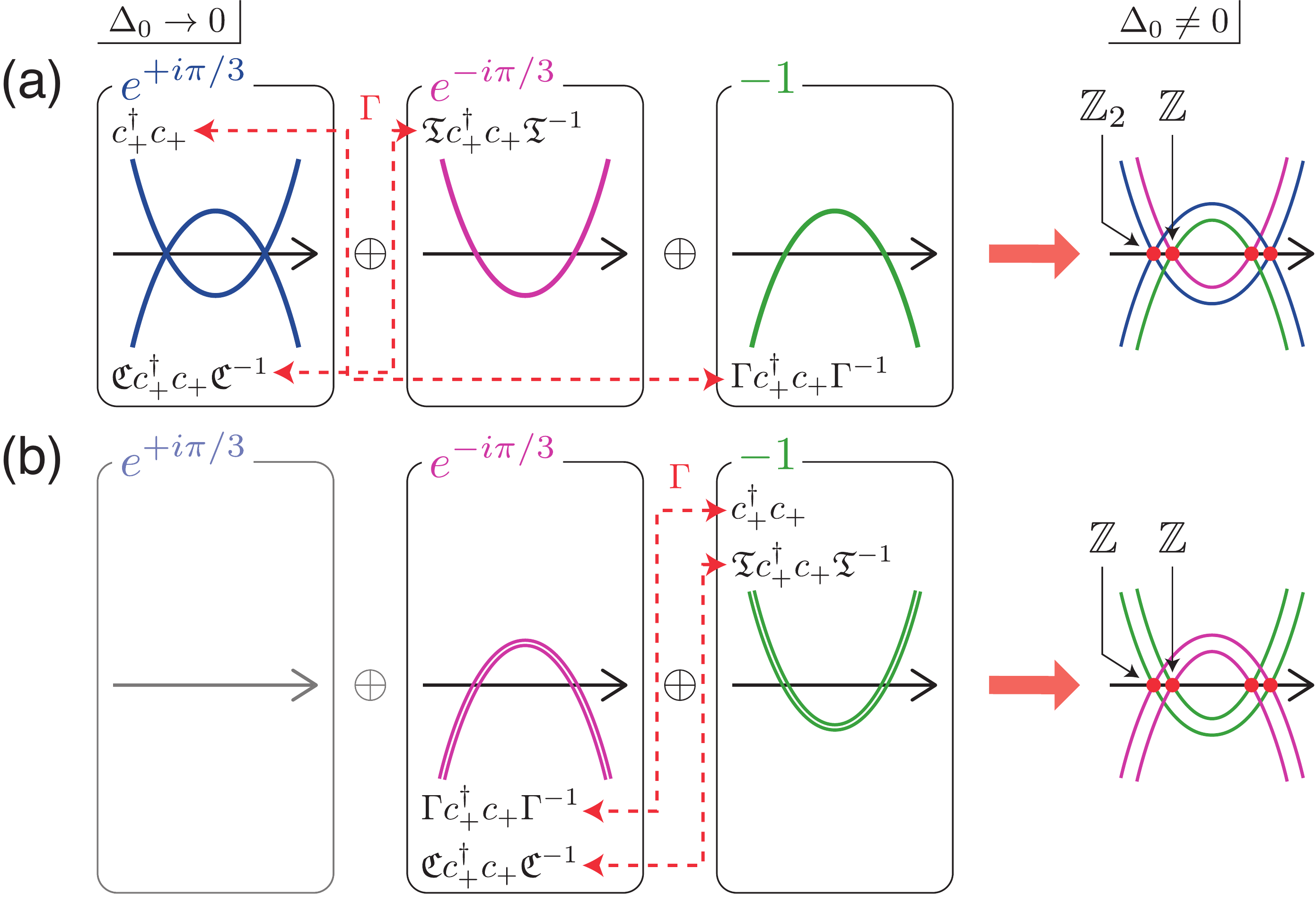}
 \caption{Band-theoretical picture of the $^{2}E_g$ symmetric BdG Hamiltonian on a $C_3$ line for (a) $\alpha = \pm 1 / 2$ and (b) $\alpha = \pm 3 / 2$ normal Bloch states. $c_+^\dagger$ in (a) and (b) is the abbreviated notation of $c_{\bm{k}, + 1 / 2}^\dagger$ and $c_{\bm{k}, + 3 / 2}^\dagger$, respectively. In the right panels, the red points represent nodes on the line. Although the particle bands and the hole bands for $\Delta_0 = 0$ are doubly degenerated, they split for a finite $\Delta_0$ due to the TRS breaking in the superconducting state.}
 \label{fig:eigenvalue_C3_Eg}
\end{figure*}

Then, we discuss the topological numbers for $\alpha = \pm 1 / 2$ bands.
Reflecting the $C_3$ symmetry, the BdG Hamiltonian matrix commutes with the threefold rotation matrix $\hat{U}_{\text{BdG}}^{C_3}$:
\begin{equation}
 [\hat{H}_{\text{BdG}}(\bm{k}), \hat{U}_{\text{BdG}}^{C_3}] = 0, \label{eq:commutation_Hamiltonian_C3}
\end{equation}
which indicates that $\hat{H}_{\text{BdG}}(\bm{k})$ and $\hat{U}_{\text{BdG}}^{C_3}$ are simultaneously block-diagonalized.
There exists a unitary matrix $\hat{V}$ such that
\begin{align}
 \hat{H}_{\text{BdG}}(\bm{k}) &= \hat{V} \begin{pmatrix}
                                          \hat{H}_{e^{+ i\pi / 3}}(\bm{k}) & 0 & 0 \\
                                          0 & \hat{H}_{e^{- i\pi / 3}}(\bm{k}) & 0 \\
                                          0 & 0 & \hat{H}_{-1}(\bm{k})
                                         \end{pmatrix} \hat{V}^\dagger, \label{eq:block-diagonalized_Hamiltonian_C3_1/2} \displaybreak[2] \\
 \hat{U}_{\text{BdG}}^{C_3} &= \hat{V} \begin{pmatrix}
                                        e^{+ i\pi / 3} \bm{1}_2 & 0 & 0 \\
                                        0 & e^{- i\pi / 3} & 0 \\
                                        0 & 0 & -1
                                       \end{pmatrix} \hat{V}^\dagger.
\end{align}
The block-diagonalized Hamiltonians $H_{e^{+ i\pi / 3}}(\bm{k})$, $H_{e^{- i\pi / 3}}(\bm{k})$, and $H_{-1}(\bm{k})$ are written as follows:
\begin{align}
 & H_{e^{+ i\pi / 3}}(\bm{k}) = \frac{1}{2} (c_{\bm{k}, + 1 / 2}^\dagger, \, \mathfrak{C} c_{\bm{k}, + 1 / 2}^\dagger \mathfrak{C}^{-1}) \notag \\
 & \quad\times
 \begin{pmatrix}
  \xi_{\bm{k}} - h_{\bm{k}} \Delta_0^2 & 0 \\
  0 & - (\xi_{\bm{k}} - h_{\bm{k}} \Delta_0^2)
 \end{pmatrix}
 \begin{pmatrix}
  c_{\bm{k}, + 1 / 2} \\
  \mathfrak{C} c_{\bm{k}, + 1 / 2} \mathfrak{C}^{-1}
 \end{pmatrix}, \displaybreak[2] \\
 & H_{e^{- i\pi / 3}}(\bm{k}) = \frac{1}{2} \mathfrak{T} c_{\bm{k}, + 1 / 2}^\dagger \mathfrak{T}^{-1} (\xi_{\bm{k}} + h_{\bm{k}} \Delta_0^2) \mathfrak{T} c_{\bm{k}, + 1 / 2} \mathfrak{T}^{-1}, \displaybreak[2] \\
 & H_{-1}(\bm{k}) = \frac{1}{2} \Gamma c_{\bm{k}, + 1 / 2}^\dagger \Gamma^{-1} (- \xi_{\bm{k}} - h_{\bm{k}} \Delta_0^2) \Gamma c_{\bm{k}, + 1 / 2} \Gamma^{-1}.
\end{align}
The band structures obtained by the BdG Hamiltonian~\eqref{eq:block-diagonalized_Hamiltonian_C3_1/2} are schematically shown in Fig.~\ref{fig:eigenvalue_C3_Eg}(a).
For the $\Delta_0 \to 0$ limit, we can identify the eigenvalue of $C_3$ for each band by using the property of $\mathfrak{T}$, $\mathfrak{C}$, and $\Gamma$.
Even when $\Delta_0$ is finite, i.e., the TRS $\mathfrak{T}$ and the CS $\Gamma$ are broken, the eigenvalues are not changed since the commutation relation Eq.~\eqref{eq:commutation_Hamiltonian_C3} remains preserved.
Then, the bands split due to the pseudomagnetic field $h_{\bm{k}} \Delta_0^2$, and they create nodes at the zero energy because the \textit{spin-singlet} $^{2}E_g$ gap function cannot have offdiagonal components in the same eigenspace.
The nodes corresponding to the $\alpha = - 1 / 2$ particle band [the pink one in Fig.~\ref{fig:eigenvalue_C3_Eg}(a)] are obviously characterized by the $\mathbb{Z}$ number,
\begin{equation}
 \nu_- \equiv \#(\text{occupied states of $\hat{H}_{e^{- i\pi / 3}}(\bm{k})$}) \in \mathbb{Z}.
\end{equation}
On the other hand, the nodes corresponding to the $\alpha = + 1 / 2$ particle band [the blue one in Fig.~\ref{fig:eigenvalue_C3_Eg}(a)] cannot be characterized by the filling, since there simultaneously exists the hole band belonging to the same eigenspace of $C_3$.
Instead, the PHS $\mathfrak{C}$ with $\mathfrak{C}^2 = + E$ in the eigenspace ensures that these nodes are protected by the $\mathbb{Z}_2$ index~\cite{Ghosh2010, Agterberg2017},
\begin{equation}
 (-1)^{l_+} \equiv \sgn[i^n \Pf\{\hat{U}_{\mathfrak{C}, e^{+ i\pi / 3}} \hat{H}_{e^{+ i\pi / 3}}(\bm{k})\}] \in \mathbb{Z}_2,
\end{equation}
with $n = \dim(\hat{H}_{e^{+ i\pi / 3}}) / 2$.

The topological numbers for $\alpha = \pm 3 / 2$ bands are obtained in the following.
In this case, the BdG Hamiltonian matrix and the threefold rotation matrix are simultaneously block-diagonalized as
\begin{align}
 \hat{H}_{\text{BdG}}(\bm{k}) &= \hat{V} \begin{pmatrix}
                                          \hat{H}_{-1}(\bm{k}) & 0 \\
                                          0 & \hat{H}_{e^{- i\pi / 3}}(\bm{k})
                                         \end{pmatrix} \hat{V}^\dagger, \label{eq:block-diagonalized_Hamiltonian_C3_3/2} \\
 \hat{U}_{\text{BdG}}^{C_3} &= \hat{V} \begin{pmatrix}
                                        -1 & 0 \\
                                        0 & e^{- i\pi / 3}
                                       \end{pmatrix} \hat{V}^\dagger,
\end{align}
where
\begin{align}
 & H_{-1}(\bm{k}) = \frac{1}{2} (c_{\bm{k}, + 3 / 2}^\dagger, \, \mathfrak{T} c_{\bm{k}, + 3 / 2}^\dagger \mathfrak{T}^{-1}) \notag \\
 & \quad\times
 \begin{pmatrix}
  \xi_{\bm{k}} - h_{\bm{k}} \Delta_0^2 & 0 \\
  0 & \xi_{\bm{k}} + h_{\bm{k}} \Delta_0^2
 \end{pmatrix}
 \begin{pmatrix}
  c_{\bm{k}, + 3 / 2} \\
  \mathfrak{T} c_{\bm{k}, + 3 / 2} \mathfrak{T}^{-1}
 \end{pmatrix}, \displaybreak[2] \\
 & H_{e^{- i\pi / 3}}(\bm{k}) = \frac{1}{2} (\mathfrak{C} c_{\bm{k}, + 3 / 2}^\dagger \mathfrak{C}^{-1}, \, \Gamma c_{\bm{k}, + 3 / 2}^\dagger \Gamma^{-1}) \notag \\
 & \quad\times
 \begin{pmatrix}
  - (\xi_{\bm{k}} - h_{\bm{k}} \Delta_0^2) & 0 \\
  0 & - (\xi_{\bm{k}} + h_{\bm{k}} \Delta_0^2)
 \end{pmatrix}
 \begin{pmatrix}
  \mathfrak{C} c_{\bm{k}, + 3 / 2} \mathfrak{C}^{-1} \\
  \Gamma c_{\bm{k}, + 3 / 2} \Gamma^{-1}
 \end{pmatrix}.
\end{align}
Figure~\ref{fig:eigenvalue_C3_Eg}(b) shows the band structures obtained by the BdG Hamiltonian matrix~\eqref{eq:block-diagonalized_Hamiltonian_C3_3/2}.
For both of the $\alpha = \pm 3 / 2$ particle bands [the green ones in Fig.~\ref{fig:eigenvalue_C3_Eg}(b)], the nodes on the $C_3$-symmetric line are characterized by the $\mathbb{Z}$ number,
\begin{equation}
 \nu_\pm \equiv \#(\text{occupied states of $\hat{H}_{- 1}(\bm{k})$}) \in \mathbb{Z}.
\end{equation}

Irrespective of the IR of the normal band $\alpha$, the TRS breaking even-parity $^{2}E_g$ order parameter induces the band splitting and nodes on the $C_3$-symmetric axis.
These facts indicate that the nodes are parts of inflated Bogoliubov FSs, which are characterized by the Pfaffian of the antisymmetrized BdG Hamiltonian, $P(\bm{k}) \in \mathbb{Z}_2$, defined for all $\bm{k}$~\cite{Agterberg2017}.
Indeed, the previous studies have suggested the existence of Bogoliubov FSs in SrPtAs~\cite{Agterberg2017, Bzdusek2017, Brydon2018}.
In addition, our theory finds that the topological protection of the nodes on the high-symmetric line is differently defined in response to the IR $\alpha$.
Although the above intuitive discussions are based on the single-band model in the weak coupling limit, the topological protection ensures the stability of nodes against multiband effects.

\subsection{Line nodes: \texorpdfstring{CeCoIn$_5$ with $B_{1g}$}{CeCoIn5 with B1g} order parameter}
\label{sec:CeCoIn5}
\subsubsection{Background}
A Ce-based heavy-fermion compound CeCoIn$_5$ is a tetragonal lattice which is characterized by the space group $P4/mmm$ ($D_{4h}^1$).
According to angle-resolved photoemission spectroscopy studies~\cite{Koitzsch2009, Koitzsch2013}, de Haas van Alphen measurements~\cite{Hall2001, Settai2001, Shishido2002, Elgazzar2004}, and first-principles calculations~\cite{Choi2012, Nomoto2014}, the compound possesses three-dimensional FSs crossing the high-symmetry $\Gamma$-$Z$ line.
CeCoIn$_5$ shows superconductivity at ambient pressure below 2.3 K~\cite{Petrovic2001}.
Regarding the pairing symmetry in this compound, the presence of line nodes on the FS is indicated by specific heat and thermal conductivity measurement~\cite{Movshovich2001}, and nuclear quadrupole resonance relaxation rate $1 / T_1$~\cite{Kohori2001}.
Furthermore, scanning tunneling spectroscopy~\cite{Akbari2011, Allan2013, Zhou2013}, field-angle-resolved measurements of thermal conductivity~\cite{Izawa2001} and heat capacity~\cite{An2010}, and torque magnetometry~\cite{Xiao2008} strongly suggest that the superconducting order parameter possesses $d_{x^2 - y^2}$-wave ($B_{1g}$) symmetry.

\subsubsection{Classification on \texorpdfstring{$\Gamma$-$Z$}{Gamma-Z} line}
We classify the gap structure on the $\Gamma$-$Z$ line with the $C_{4v}$ symmetry.
According to the group-theoretical classification, the gap structure on the line is nodal for the $d_{x^2 - y^2}$-wave order parameter belonging to the $B_{1g}$ IR of $D_{4h}$, irrespective of the angular momentum of the normal Bloch state (Table~\ref{tab:gap_classification_group_theory}).
The nodes on the axis are a part of a line node because of the following reason.
In the BZ of $D_{4h}$, the $\Gamma$-$Z$-$A$-$M$ plane is invariant under the diagonal mirror $\sigma_{[1 1 0]}$.
Therefore the combination of the compatibility relation
\begin{equation}
 (B_{1g} \ \text{of} \ D_{4h}) \downarrow C_{2h}^{[1 1 0]} = B_g,
\end{equation}
and the gap classification on the plane~\cite{Micklitz2017_PRL, Sumita2018}
\begin{equation}
 P^{\bm{k}} = A_g + 2 A_u + B_u,
\end{equation}
shows that a line node emerge on the $[1 1 0]$ plane.\footnote{The same result is obtained for the $[1 {-1} 0]$ plane.}
The line node is protected by the 1D winding number $W_l \in \mathbb{Z}$ defined on a loop $l$ encircling the nodal line, using the CS $\Gamma$~\cite{Kobayashi2014, Kobayashi2018}.
Thus the node is stable even when the mirror symmetry $\sigma_{[1 1 0]}$ is broken as long as the CS exists.
More generally, when the superconducting order parameter is even-parity and preserves TRS, nodes on a high-symmetry axis may be a part of (trivial) line nodes [see (L) in Table~\ref{tab:gap_classification_topological}].

Table~\ref{tab:gap_classification_topological}(g) shows that the topological classification for the $B_{1g}$ order parameter is $\mathbb{Z}$.
The orthogonality test for the CS $\Gamma$ is $W_\alpha^\Gamma = 0$, which indicates that $\Gamma$ changes the bases of the IR $\alpha = 1 / 2$ ($3 / 2$) to those of $\alpha' = 3 / 2$ ($1 / 2$).
Therefore the $\mathbb{Z}$ index is defined by the number of occupied states of the block-diagonalized Hamiltonian belonging to the IR $\alpha$.

\subsection{Point nodes: \texorpdfstring{UPt$_3$ with $E_{2u}$}{UPt3 with E2u} order parameter}
\label{sec:UPt3}
\subsubsection{Background}
\begin{figure}[tbp]
 \centering
 \includegraphics[width=6cm, clip]{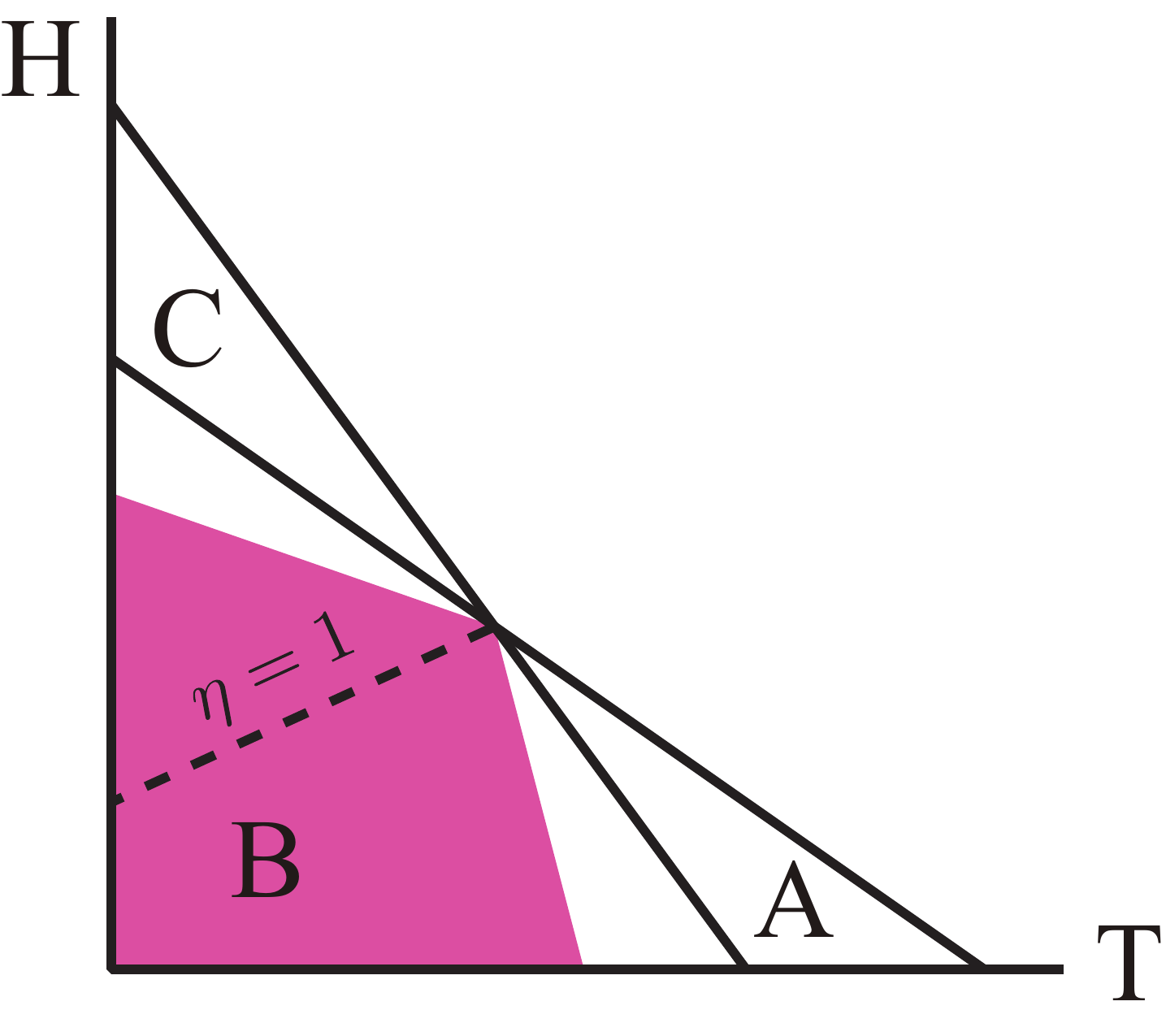}
 \caption{Multiple superconducting phases of UPt$_3$ in the magnetic field-temperature plane~\cite{Bruls1990, Adenwalla1990, Sauls1994, Tou1998, Joynt2002}. The shaded region shows the Weyl superconducting phase~\cite{Yanase2016, Yanase2017}. In the system, $C_3$ symmetry is preserved only on the dashed line.}
 \label{fig:UPt3_phase}
\end{figure}

UPt$_3$ is a heavy-fermion superconductor which has been intensively investigated after the discovery of superconductivity in 1980s~\cite{Stewart1984}.
The crystal symmetry of UPt$_3$ is represented by the space group $P6_3/mmc$ ($D_{6h}^4$),\footnote{Symmetry breaking by a weak crystal distortion has been reported~\cite{Walko2001}, although its reliability is under debate. We here assume high-symmetry space group $P6_3/mmc$.} which is the same group as SrPtAs.
Multiple superconducting phases illustrated in Fig.~\ref{fig:UPt3_phase}~\cite{Fisher1989, Bruls1990, Adenwalla1990, Sauls1994, Tou1998, Joynt2002} unambiguously exhibit exotic Cooper pairing which is probably categorized into the 2D IR of point group $D_{6h}$~\cite{Sigrist-Ueda}.
After several theoretical proposals examined by experiments for more than three decades, the $E_{2u}$ representation has been regarded as the most reasonable symmetry of superconducting order parameter~\cite{Sauls1994, Joynt2002}.
In particular, the multiple superconducting phases in the temperature--magnetic-field plane are naturally reproduced by assuming a weak symmetry-breaking term of hexagonal symmetry~\cite{Sauls1994}.
Furthermore, a phase-sensitive measurement~\cite{Strand2010} and the observation of spontaneous TRS breaking~\cite{Luke1993, Schemm2014} in the low-temperature and low-magnetic-field B phase, which was predicted in the $E_{2u}$ state, support the $E_{2u}$ symmetry of superconductivity.

Recent our studies have investigated nontrivial gap structures in UPt$_3$ by using the effective two-sublattice single-orbital model~\cite{Yanase2016, Kobayashi2016, Yanase2017, Sumita2018}.
Especially on the $C_{3v}$-symmetric $K$-$H$ line, we have reported that the existence of point nodes depends on the angular momentum of the normal Bloch state by taking into account the inter-sublattice $p$-wave component~\cite{Sumita2018}.
In the model, the two-component superconducting order parameter in the $E_{2u}$ IR is considered:
\begin{equation}
 \hat{\Delta}(\bm{k}) = \eta_1 \hat{\Gamma}_1^{E_{2u}} + \eta_2 \hat{\Gamma}_2^{E_{2u}}, \label{eq:UPt3_order}
\end{equation}
which are parametrized as
\begin{equation}
 (\eta_1, \eta_2) = \Delta (1, i\eta) / \sqrt{1 + \eta^2}, \label{eq:UPt3_eta}
\end{equation}
with a real variable $\eta$.
Note that $\hat{\Gamma}_1^{E_{2u}}$ and $\hat{\Gamma}_2^{E_{2u}}$ are not equivalent to rotation-invariant bases of $C_3$ [see Eq.~\eqref{eq:C3_order_pm}].

The A, B, and C superconducting phases illustrated in Fig.~\ref{fig:UPt3_phase}~\cite{Fisher1989, Bruls1990, Adenwalla1990, Sauls1994, Tou1998, Joynt2002} are characterized by the ratio of two-component order parameters $\eta = \eta_2 / i \eta_1$ summarized in Table~\ref{tab:range_eta}.
A pure imaginary ratio of $\eta_1$ and $\eta_2$ in the B phase implies the chiral superconducting state, which maximally gains the condensation energy.
Owing to the $p$-wave components, the B phase is a nonunitary state.
Indeed, a recent theoretical study based on our model~\cite{WangZ2017, Triola2018} has shown the polar Kerr effect consistent with the experiment~\cite{Schemm2014}.

\begin{table}[tbp]
 \caption{Range of the parameter $\eta$ in the A, B, and C phases of UPt$_3$.}
 \label{tab:range_eta}
 \begin{center}
  \begin{tabular}{cc} \hline\hline
   A phase & $|\eta| = \infty$ \\
   B phase & $0 < |\eta| < \infty$ \\
   C phase & $|\eta| = 0$ \\ \hline\hline
  \end{tabular}
 \end{center}
\end{table}

\subsubsection{Classification on \texorpdfstring{$K$-$H$}{K-H} line}
Here we clarify the gap structure on the $K$-$H$ line by symmetry~\cite{Sumita2018} and topology.
As is the case for SrPtAs, we consider the compatibility relation: $(E_{2u} \ \text{of} \ D_{6h}) \downarrow D_{3d} = E_u$.
The group-theoretical classification elucidates that the gap structure on the $K$-$H$ line with the $C_{3v}$ symmetry is fully gapped (nodal) when the normal Bloch state is $E_{1 / 2}$ ($E_{3 / 2}$) with the $E_u$ order parameter (Table~\ref{tab:gap_classification_group_theory}).
This is an important example of the $j_z$-dependent gap structure suggested in the previous study~\cite{Sumita2018}.
From the viewpoint of topology, Tables~\ref{tab:gap_classification_topological}(b1) and \ref{tab:gap_classification_topological}(b2) shows that the topological classifications for an $^{1, 2}E_u$ order parameter indeed depend on the angular momentum: $0 \oplus \mathbb{Z}$ for $\alpha = \pm 1 / 2$ and $\mathbb{Z} \oplus \mathbb{Z}$ for $\alpha = \pm 3 / 2$.
However, in this case, the topological classification is slightly different from the gap classification by symmetry.
For one of the $\alpha = \pm 1 / 2$ states, the gap opens while the gap closes for the other state.
Which one is gapped depends on the $C_3$ eigenvalue of the order parameter.
The inconsistency is due to the fact that the TRS breaking is taken into account in the topological argument although it is not in the symmetry analysis.
In this case the topological classification predicts a correct gap structure.

\begin{figure}[tbp]
 \centering
 \includegraphics[width=8cm, clip]{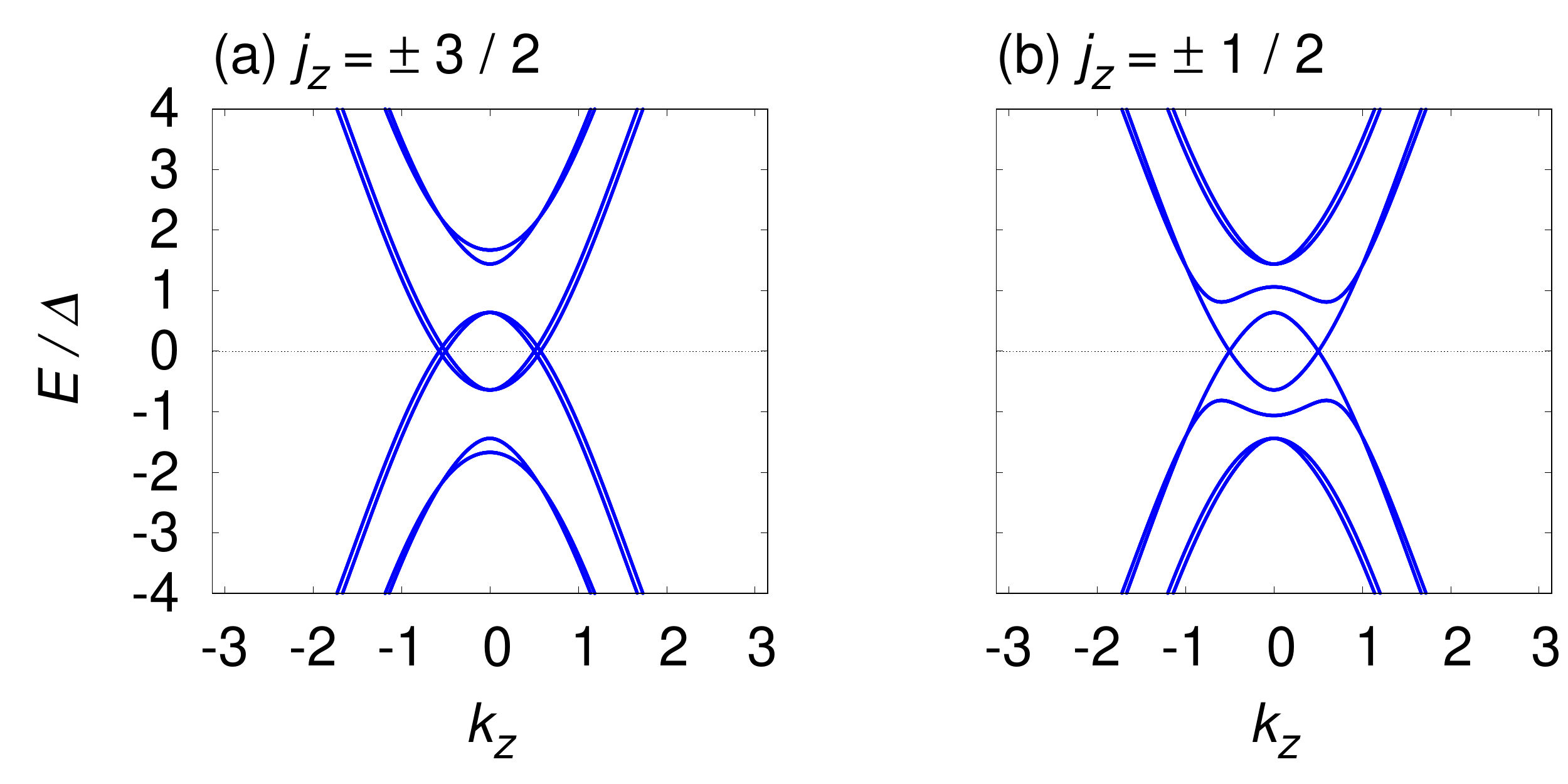}
 \caption{The quasiparticle energy dispersion on the $K$-$H$ line for (a) $j_z = \pm 3 / 2$ bands and (b) $j_z = \pm 1 / 2$ bands obtained by the model in Ref.~\cite{Sumita2018}. We assume the $C_3$ preserving and TRS breaking B phase ($\eta = 1$). The other parameters $(t, t_z, t', \mu, \Delta, \delta_1, \delta_2) = (1, - 1, 0.4, - 5.2, 0.5, 0.04, 0.2)$ are assumed so that the $K$-FSs of UPt$_3$ are reproduced.}
 \label{fig:UPt3_eigenKH_eta1}
\end{figure}

We demonstrate such unusual gap structures by using the effective model (see Sec.~IV~A in Ref.~\cite{Sumita2018} for details).
The TRS breaking and $C_3$ preserving order parameter ($^{1}E_u$ or $^{2}E_u$) is realized for $|\eta| = 1$ (see the dashed line in Fig.~\ref{fig:UPt3_phase}), where a topological phase transition occurs~\cite{Yanase2017}.
Therefore we diagonalize the BdG Hamiltonian for $\eta = 1$, which results in the quasiparticle energy dispersion in Fig.~\ref{fig:UPt3_eigenKH_eta1}.
When the normal bands belong to $\alpha = \pm 3 / 2$, both bands create nodes at the zero energy [Fig.~\ref{fig:UPt3_eigenKH_eta1}(a)].
For $\alpha = \pm 1 / 2$, on the other hand, one band holds nodes but the other one is fully gapped [Fig.~\ref{fig:UPt3_eigenKH_eta1}(b)].
In both IRs, the nodes on the $K$-$H$ line are topologically characterized by the number of occupied states of the block-diagonalized Hamiltonian, which is obtained by using the threefold rotation matrix $\hat{U}_{\text{BdG}}^{C_3}$ in a similar way to the case of SrPtAs.
Note that Ref.~\cite{Sumita2018} showed the results for $\eta = 0, \infty$, but $\eta = 1$ has not been considered.

\subsubsection{Application to Weyl superconductivity in B phase}
\begin{figure}[tbp]
 \centering
 \includegraphics[width=8cm, clip]{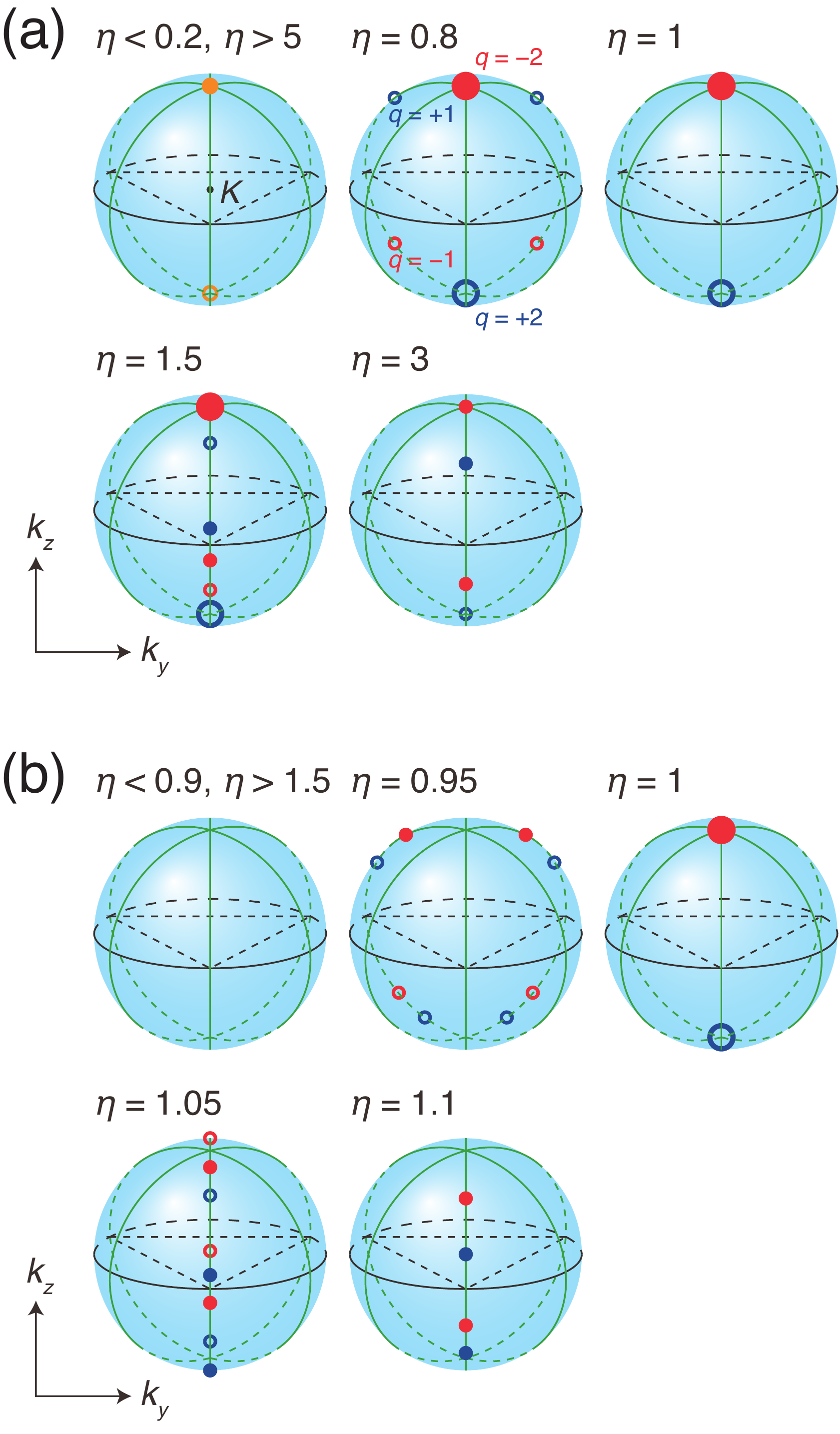}
 \caption{Illustration of pair creation and annihilation of Weyl nodes on the $K$-FS for (a) $j_z = \pm 3 / 2$ bands and (b) $j_z = \pm 1 / 2$ bands. Blue and red circles show single Weyl nodes with $q_i = 1$ and $- 1$, respectively. Large circles are double Weyl nodes with $q_i = \pm 2$. Orange circles are trivial point nodes protected by $C_3$ symmetry. Closed (open) circles represent nodes on the front (back) side of the FS.}
 \label{fig:UPt3_Weyl_node}
\end{figure}

\begin{figure*}[tbp]
 \centering
 \includegraphics[width=16cm, clip]{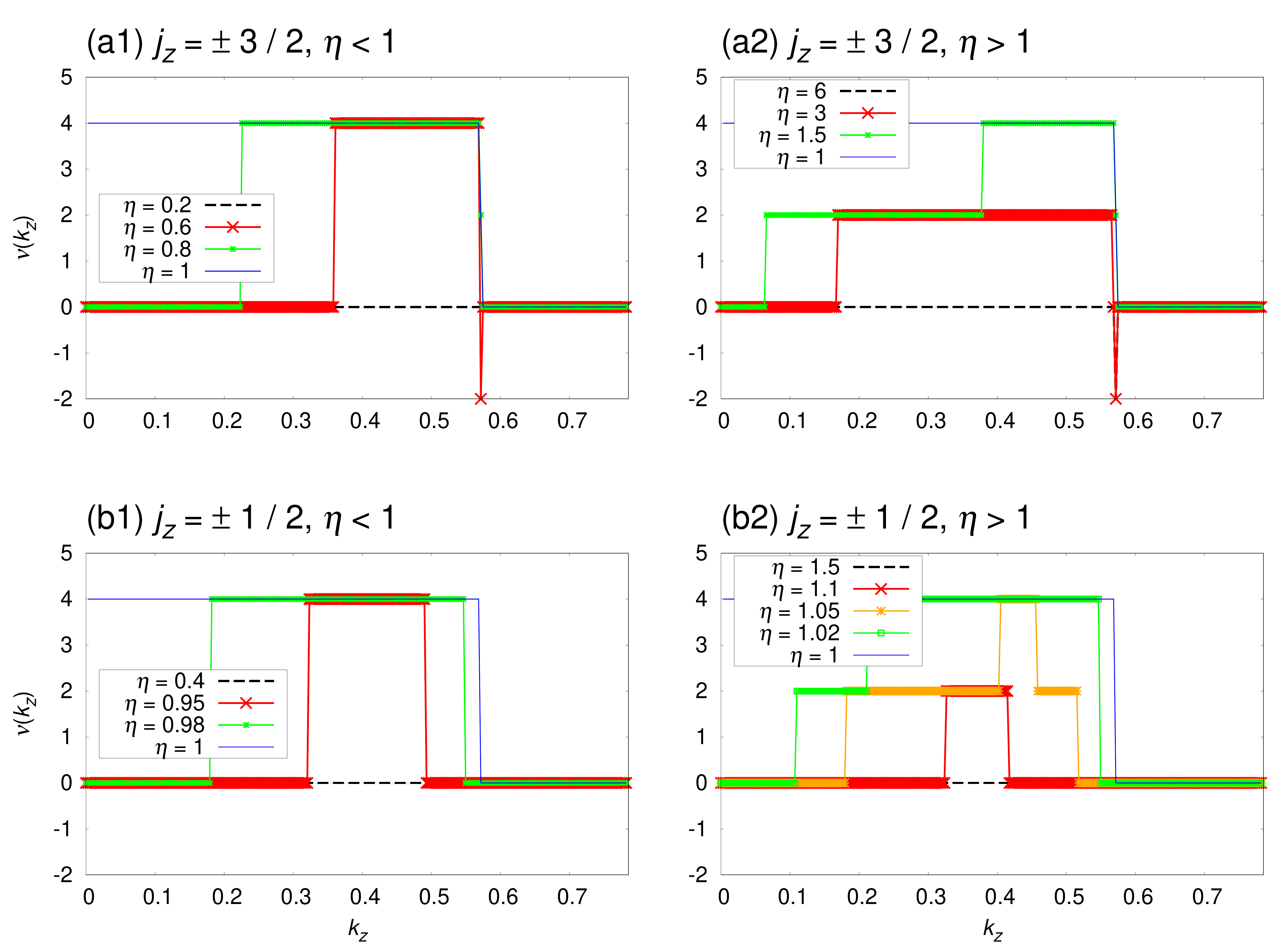}
 \caption{Chern number of the 2D BdG Hamiltonian parametrized by $k_z$ for (a) $j_z = \pm 3 / 2$ bands and (b) $j_z = \pm 1 / 2$ bands on the $K$-FS reproduced by the parameter set $(t, t_z, t', \alpha, \mu, \Delta, \delta) = (1, -1, 0.4, 0.2, -5.2, 0.1, 0.04)$.}
 \label{fig:UPt3_Chern_K-FS}
\end{figure*}

We also calculate the nodal structure in the B phase for a general parameter $0 < \eta < \infty$.
In the TRS breaking phase, there generally exists Weyl nodes characterized by a topological Weyl charge, which is defined by a monopole of Berry flux,
\begin{equation}
 q_i = \frac{1}{2 \pi} \oint_S d\bm{k} \, \bm{F}(\bm{k}).
\end{equation}
Here, the Berry flux
\begin{equation}
 F_i(\bm{k}) = - i \epsilon^{ijk} \sum_{E_n(\bm{k}) < 0} \partial_{k_j} \braket{u_n(\bm{k}) | \partial_{k_k} u_n(\bm{k})},
\end{equation}
is integrated on a closed surface $S$ surrounding an isolated point node.
We identify Weyl nodes by calculating $k_z$-dependent Chern number,
\begin{equation}
 \nu(k_z) = \frac{1}{2 \pi} \int dk_x dk_y \, F_z(\bm{k}),
\end{equation}
on a 2D $k_x$-$k_y$ plane~\cite{Thouless1982, Kohmoto1985, Fukui2005}.
An $n$-th wave function and energy of Bogoliubov quasiparticles are denoted by $\ket{u_n(\bm{k})}$ and $E_n(\bm{k})$, respectively.
By definition, when the Chern number jumps at $k_z$, its value is equal to the sum of Weyl charges at $k_z$:
\begin{equation}
 \nu(k_z + 0) - \nu(k_z - 0) = \sum_{i} q_i.
\end{equation}
Therefore we can identify Weyl charges by counting point nodes and comparing it with a jump in $\nu(k_z)$.
Indeed, the previous study has reported the presence of many Weyl nodes on the $\Gamma$- and $A$-FSs~\cite{Yanase2016}.

By using the above method, we obtain the superconducting gap structures on the $K$-FS illustrated in Fig.~\ref{fig:UPt3_Weyl_node}.
Weyl nodes (blue and red circles in Fig.~\ref{fig:UPt3_Weyl_node}) appear in the B phase, in addition to the symmetry-protected point nodes for the $j_z = \pm 3 / 2$ normal Bloch state [orange circles in Fig.~\ref{fig:UPt3_Weyl_node}(a)].
The former Weyl nodes are identified by jumps of the Chern number shown in Fig.~\ref{fig:UPt3_Chern_K-FS}.
Figure~\ref{fig:UPt3_Chern_K-FS}(b1) shows that the Chern number jumps by $\pm 4$, and we find two point nodes on the $K$-FS at a certain $k_z$ [Fig.~\ref{fig:UPt3_Weyl_node}(b), $\eta = 0.95$].
$C_6$ symmetry ensures that there also exists two point nodes on the $K'$-FS at the same $k_z$.
Therefore the point nodes are identified as single Weyl nodes with a unit charge $q_i = \pm 1$.
Through the above discussion, we determine the charge of Weyl nodes as depicted in Fig.~\ref{fig:UPt3_Weyl_node}.

It is shown that the position of Weyl nodes changes as a function of $\eta$, which indicates that the nodal structure significantly depends on the temperature and magnetic field in UPt$_3$.
For $j_z = \pm 3 / 2$ bands, the A and C phases host symmetry-protected point nodes on the north and south pole of the FS [Fig.~\ref{fig:UPt3_Weyl_node}(a), $\eta < 0.2$ and $\eta > 5$].
With an influence from the gap zeros, Weyl nodes appear in the comparatively wide range of the B phase: $0.8 \lesssim \eta \lesssim 3$.
On the other hand, there is no node in the A and C phases for $j_z = \pm 1 / 2$ bands [Fig.~\ref{fig:UPt3_Weyl_node}(b), $\eta < 0.9$ and $\eta > 1.5$].
Therefore the B phase hosts Weyl nodes only in the narrow region around the $\eta = 1$ line in Fig.~\ref{fig:UPt3_phase}: $0.95 \lesssim \eta \lesssim 1.1$.
As a result, reflecting the angular momentum dependence of the gap classification on the $C_{3v}$-symmetric $K$-$H$ line, the structures of Weyl nodes are obviously different between $j_z = \pm 3 / 2$ [Fig.~\ref{fig:UPt3_Weyl_node}(a)] and $j_z = \pm 1 / 2$ [Fig.~\ref{fig:UPt3_Weyl_node}(b)].
The nodal structure of Weyl superconductors may be clarified by the thermal Hall conductivity~\cite{Sumiyoshi2013}.

\subsection{Point nodes: UCoGe with \texorpdfstring{$B_{1u}$}{B1u} order parameter}
\label{sec:UCoGe}
\subsubsection{Background}
UCoGe is a orthorhombic superconductor whose crystal structure belongs to the nonsymmorphic space group $Pnma$ ($D_{2h}^{16}$)~\cite{Canepa1996}.
In this material, superconductivity at ambient pressure coexists with ferromagnetism~\cite{Huy2007, Aoki2014}, and therefore, odd-parity superconductivity is strongly suggested.
The high-pressure superconducting phase~\cite{Huy2007, Aoki2014, Hassinger2008, Slooten2009, Bastien2016, Cheung2016, Mineev2017} ($S_2$ phase in Ref.~\cite{Slooten2009}), where TRS recovers by the vanishing ferromagnetic moments~\cite{Hassinger2008, Slooten2009, Bastien2016, Manago2018}, is also expected to be odd-parity superconductivity, since it is continuously connected to the ferromagnetic superconducting phase~\cite{Hassinger2008, Slooten2009, Bastien2016, Cheung2016, Mineev2017}.

Although the symmetry of superconductivity in UCoGe is still under debate, recent studies have suggested the possibility of topological crystalline superconductivity, which is characterized by $\mathbb{Z}_4$ and/or $\mathbb{Z}_2$ indices, for all odd-parity IRs ($A_u$, $B_{1u}$, $B_{2u}$, and $B_{3u}$) of the superconducting order parameter~\cite{Daido2018_arXiv, Yoshida2018_arXiv}.
Especially, when the order parameter belongs to the $B_{1u}$ IR of $D_{2h}$, point nodes emerge on the $\Gamma$-$Z$ line~\cite{Yoshida2018_arXiv}.
In the following discussion, therefore, we focus on the $B_{1u}$ order parameter as a candidate of the odd-parity superconducting order.

\subsubsection{Classification on \texorpdfstring{$\Gamma$-$Z$}{Gamma-Z} line}
Considering the paramagnetic high-pressure superconducting phase, we classify the gap structure on the $\Gamma$-$Z$ line which has $C_{2v}$ symmetry.
Since the group-theoretical classification of $C_{2v}$-symmetric line is $P^{\bm{k}} = A_g + A_u + B_{2u} + B_{3u}$ (Table~\ref{tab:gap_classification_group_theory}), point nodes appear on the line in the $B_{1u}$ superconducting state.
Note that the gap classification results on the other eleven $C_{2v}$-symmetric lines in the BZ are changed due to the nonsymmorphic symmetry of $Pnma$~\cite{Yoshida2018_arXiv}.
According to Table~\ref{tab:gap_classification_topological}(e), the point nodes are characterized by the $\mathbb{Z}_2$ topological number.

Now we identify the $\mathbb{Z}_2$ topological number.
First, the IR of the little group $C_{2v}$ is written by
\begin{subequations}
 \label{eq:IR_C2v}
 \begin{align}
  \bar{\gamma}^{\bm{k}}_{1 / 2}(E) &= \hat{\sigma}_0, \\
  \bar{\gamma}^{\bm{k}}_{1 / 2}(C_2) &= - i\hat{\sigma}_z, \\
  \bar{\gamma}^{\bm{k}}_{1 / 2}(\sigma_y) &= - i\hat{\sigma}_y, \\
  \bar{\gamma}^{\bm{k}}_{1 / 2}(\sigma_x) &= - i\hat{\sigma}_x,
 \end{align}
\end{subequations}
where the bases ($c_{\bm{k}, \pm i}^\dagger$) are the eigenstates of $C_2$ with the eigenvalues $\pm i$.
Thus the minimal BdG Hamiltonian is expressed by the four bases, $c_{\bm{k}, \pm i}^\dagger$ and $\mathfrak{C} c_{\bm{k}, \pm i}^\dagger \mathfrak{C}^{-1}$.
Here, we remark that the PHS $\mathfrak{C}$ changes the eigenvalue of $C_2$ since $\mathfrak{C}$ is an antiunitary operator with $[\mathfrak{C}, C_2] = 0$ in the $B_{1u}$ superconducting state.
Furthermore, Eq.~\eqref{eq:IR_C2v} shows that the mirror operator $\sigma_y$ also changes the eigenvalue of $C_2$.
Therefore a new PHS operator $\widetilde{\mathfrak{C}} \equiv \mathfrak{C} \sigma_y$ preserves the eigenvalue of $C_2$, and it has the following relation,
\begin{equation}
 \widetilde{\mathfrak{C}}^2 = \mathfrak{C}^2 \sigma_y^2 = + E,
\end{equation}
where we use $[\mathfrak{C}, \sigma_y] = 0$ for the $B_{1u}$ order parameter.
As a result the BdG Hamiltonian matrix $\hat{H}_{\text{BdG}}(\bm{k})$ are decomposed by $C_2$ eigen-sectors $\hat{H}_\pm(\bm{k})$, each of which has the PHS-like operator $\widetilde{\mathfrak{C}}$.
Due to the symmetry, the $\mathbb{Z}_2$ number can be defined in each sector~\cite{Ghosh2010, Agterberg2017}:
\begin{equation}
 (-1)^{l_\pm} \equiv \sgn[i^n \Pf\{\hat{U}_{\widetilde{\mathfrak{C}}, \pm} \hat{H}_\pm(\bm{k})\}] \in \mathbb{Z}_2, \label{eq:UCoGe_Z2_number}
\end{equation}
with $n = \dim(\hat{H}_\pm) / 2$.

\begin{figure}[tbp]
 \centering
 \includegraphics[width=8cm, clip]{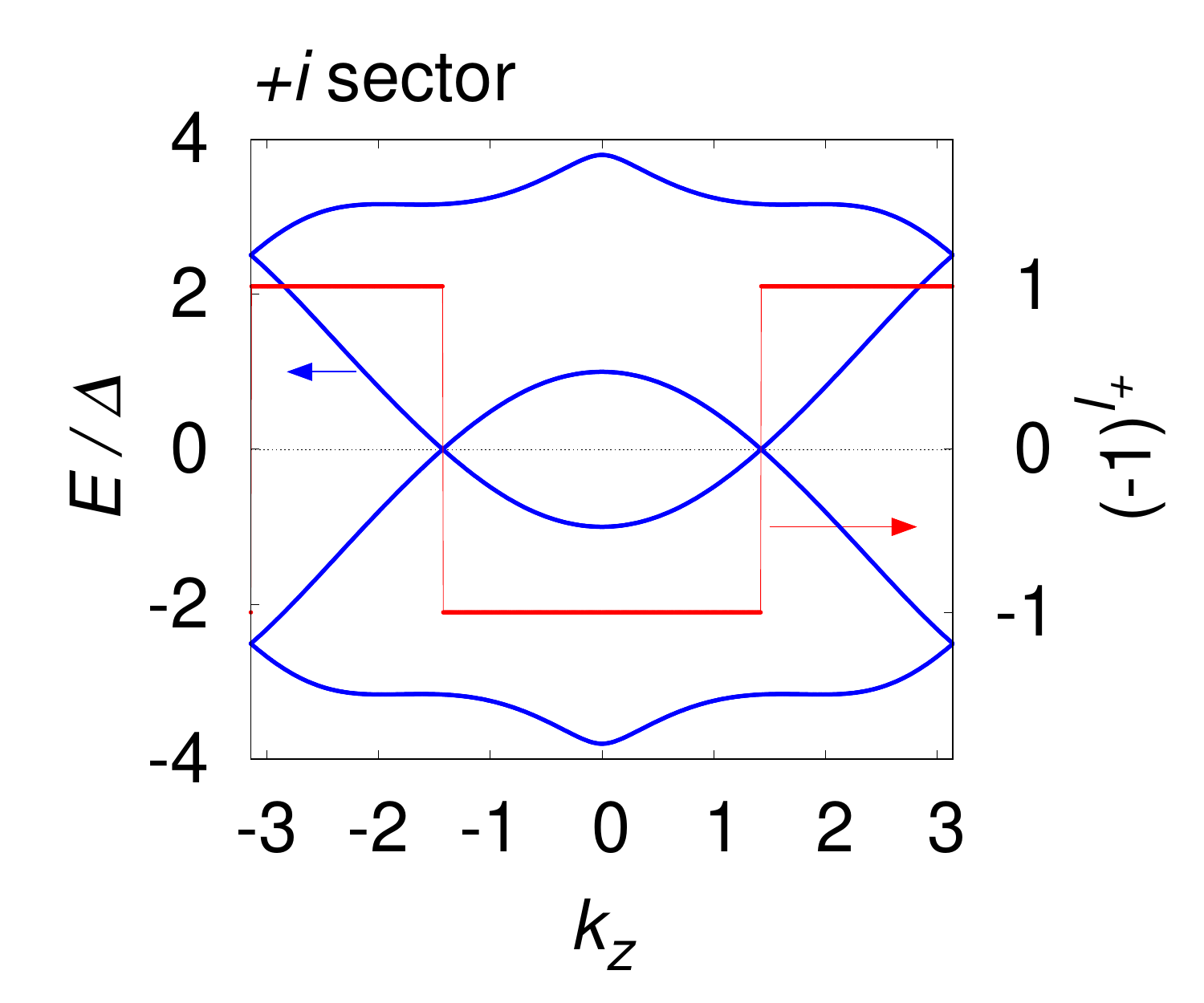}
 \caption{The quasiparticle energy dispersion (blue lines) and the $\mathbb{Z}_2$ topological number (red line) on the $\Gamma$-$Z$ line for the $+ i$ sector of the block-diagonalized BdG Hamiltonian. The result for the $- i$ sector is the same as the $+ i$ case. We adopt the effective model in Refs.~\cite{Daido2018_arXiv, Yoshida2018_arXiv}. The parameters $(t_1, t_2, t_3, t_{ab}, t_{ab}', \mu, t_1', \alpha) = (1, 0.2, 0.1, 0.5, 0.1, -0.5, 0.1, 0.3)$ are assumed so that the $\Gamma$-FSs of UCoGe are reproduced.}
 \label{fig:UCoGe_eigen_Pf}
\end{figure}

We demonstrate the topological crystalline point nodes by using the effective four-sublattice single-orbital model introduced in Refs.~\cite{Daido2018_arXiv, Yoshida2018_arXiv}.
Figure~\ref{fig:UCoGe_eigen_Pf} shows the eigenvalues of the BdG Hamiltonian and the $\mathbb{Z}_2$ index for the $+ i$ eigen-sector on the $\Gamma$-$Z$ line.
Obviously, we find that the $\mathbb{Z}_2$ topological number $(-1)^{l_+}$ changes at the point nodes.

\section{Summary and discussion}
\label{sec:summary}
In this paper, on the basis of the symmetry and topology we classified superconducting gap structure on high-symmetry $n$-fold lines in the BZ.
First, we reviewed the group-theoretical analysis using the Mackey-Bradley theorem.
On threefold and sixfold axes, the gap classification depends on the total angular momentum of the normal Bloch state $j_z$, while that is unique on twofold and fourfold axes.

Next, fixing an IR of the order parameter, we investigated symmetry of the BdG Hamiltonian on the lines by using the Wigner criteria for TRS and PHS, and the orthogonality test for CS.
The result of the tests corresponds to the EAZ class of the IR, which informs us the presence of a topological number according to the knowledge of $K$ theory.
As a result, the topological analysis completely corresponds with the above group-theoretical classification; all nodes shown by group theory are characterized by a 0D $\mathbb{Z}$ or $\mathbb{Z}_2$ index.
Thus the symmetry-protected nodes are topologically protected.
Such \textit{topological crystalline superconducting nodes} on the high-symmetry line are classified into three types: point nodes, a part of line nodes, and a part of surface nodes (Bogoliubov FSs).

Furthermore, we applied such classification to some candidate superconductors: SrPtAs (a part of surface nodes), CeCoIn$_5$ (a part of line nodes), UPt$_3$ ($j_z$-dependent point nodes), and UCoGe ($j_z$-independent point nodes).
For all cases, a $\mathbb{Z}$ index is the number of occupied states belonging to a certain eigenspace of rotation, and a $\mathbb{Z}_2$ index is the sign of the Pfaffian of antisymmetrized Hamiltonian.
In addition, we showed that the structure of Weyl nodes also depends on the angular momentum of the normal Bloch state in the TRS breaking B phase of UPt$_3$, reflecting the $j_z$-dependent gap structure on the $C_{3v}$-symmetric $K$-$H$ line.

The group-theoretical analysis helps us to easily search symmetry-protected superconducting nodes.
On the other hand, the property of topological number tells us the stability of such nodes.
Therefore it is important to complementarily classify superconducting gap structures from both aspects of group theory and topology.
Our study motivates the research community to reacknowledge the importance of such complementary studies, and to expect that all crystal symmetry-protected nodes are protected by topology.

\begin{acknowledgments}
 The authors are grateful to M.~H.~Fischer, T.~Yoshida, and A.~Daido for fruitful discussions.
 K.S. was supported by PRESTO, JST (JPMJPR18L4).
 This work was supported by Grant-in Aid for Scientific Research on Innovative Areas ``J-Physics'' (JP15H05884) and ``Topological Materials Science'' (JP16H00991 and JP18H04225) from JSPS of Japan, by ``J-Physics: Young Researchers Exchange Program'' (JP15K21732), and by JSPS KAKENHI Grants No. JP15K05164, No. JP15H05745, No. JP17J09908, No. JP18H05227, and No. JP18H01178.
\end{acknowledgments}

\appendix
\section{Wigner criterion}
\label{sec:Wigner_criterion}
We review the method and meaning of the Wigner criterion used in Secs.~\ref{sec:preparation} and \ref{sec:gap_classification_topological}.
For detailed proofs of the criterion, see Refs.~\cite{Inui-Tanabe-Onodera, Shiozaki2018_arXiv}.

\subsection{Formulation}
We first suppose that $H$ is a unitary group and $\alpha$ is a certain $d_\alpha$-dimensional IR of $H$, which has the basis functions $\psi_i$ ($i = 1, 2, \dots, d_\alpha$).
$\psi_i$ transforms under the symmetry operation $h \in H$ as
\begin{equation}
 h \psi_i = \sum_{j = 1}^{d_\alpha} \psi_j [D_\alpha(h)]_{ji},
\end{equation}
where $D_\alpha$ is a representation matrix of the IR $\alpha$.
Then, we consider whether the degeneracy of the representation increases or not by adding an antiunitary operator $a$ to the group: $H + a H$.
The problem can be solved by the Wigner criterion~\cite{Inui-Tanabe-Onodera, Shiozaki2018_arXiv}:
\begin{equation}
 W_\alpha^a \equiv \frac{1}{|H|} \sum_{h \in H} z_{a h, a h} \chi[D_\alpha((a h)^2)] =
  \begin{cases}
   1 & \text{(a),} \\
   -1 & \text{(b),} \\
   0 & \text{(c).}
  \end{cases}
\end{equation}
Here, $\{z_{h_1, h_2}\} \in Z^2(H + a H, U(1)_\phi)$ is a factor system arising in a representation
\begin{equation}
 z_{h_1, h_2} U_{h_1 h_2} = 
  \begin{cases}
   U_{h_1} U_{h_2} & (\phi(h_1) = 1), \\
   U_{h_1} U_{h_2}^* & (\phi(h_1) = -1),
  \end{cases}
\end{equation}
where $\phi: H + a H \to \mathbb{Z}_2 = \{\pm 1\}$ is an indicator for unitary/antiunitary symmetry.
The meanings of the cases (a), (b), and (c) are shown in the following.
\begin{enumerate}
 \item There is no additional degeneracy due to the presence of the antiunitary operator $a$, because $\{\psi_i\}$ and $\{a \psi_i\}$ are not independent.
 \item The presence of the operator $a$ gives rise to additional degeneracy, because $\{a \psi_i\}$ is linearly independent of $\{\psi_i\}$ although they belong to the same IR $\alpha$.
 \item The degeneracy is doubled by applying $a$, because the basis $\{a \psi_i\}$ belongs to a representation $\alpha'$ inequivalent to $\alpha$.
\end{enumerate}

\subsection{Example: \texorpdfstring{$C_3$}{C3} symmetry}
As an example of the Wigner criterion, we see the rotational property of the basis of spin angular momentum $\psi = \ket{s_z}$.
In a continuous space, $\ket{s_z}$ transforms under the $\theta$ rotation $C(\theta)$ around the $z$ axis as
\begin{equation}
 C(\theta) \ket{s_z} = e^{i \theta s_z} \ket{s_z}. \label{eq:rotation_sz}
\end{equation}
Next, we discuss threefold-rotational symmetric system $H = \{E, C_3, (C_3)^2\}$.
In this symmetry, the continuous rotational symmetry $C(\theta)$ is restricted to discrete symmetries: $\theta = 0$, $2\pi / 3$, and $4\pi / 3$.
Now we investigate the additional degeneracy by imposing the TRS ${\cal T}$ on the system for $s_z = + 3 / 2$ and $+ 1 / 2$.
\begin{enumerate}
 \setcounter{enumi}{1}
 \item $s_z = + 3 / 2$ case. \\
       The eigenvalues $e^{i \theta s_z}$ in Eq.~\eqref{eq:rotation_sz}, namely the 1D IR matrices $D_{+ 3 / 2}$ with the basis $\ket{+ \frac{3}{2}}$, are given by
       \begin{subequations}
        \begin{align}
         & D_{+ 3 / 2}(E) = 1, \\
         & D_{+ 3 / 2}(C_3) = - 1, \\
         & D_{+ 3 / 2}((C_3)^2) = 1.
        \end{align}
       \end{subequations}
       Then, by adding the TRS ${\cal T}$, the Wigner criterion leads to
       \begin{align}
        W_{+ 3 / 2}^{\cal T} &= \frac{1}{3} \sum_{h \in H} z_{{\cal T} h, {\cal T} h} \chi[D_{+ 3 / 2}(({\cal T} h)^2)] \notag \\
        &= \frac{1}{3} ( - 1 - 1 - 1 ) = -1. \label{eq:Wigner_criterion_+3/2}
       \end{align}
       Note that ${\cal T}$ commutes with all operators in $H$, and ${\cal T}^2 = (C_3)^3 = - E$.
       Eq.~\eqref{eq:Wigner_criterion_+3/2} indicates that ${\cal T} \ket{+ \frac{3}{2}} \propto \ket{- \frac{3}{2}}$ has the same rotational property as $\ket{+ \frac{3}{2}}$:
       \begin{subequations}
        \begin{align}
         & D_{- 3 / 2}(E) = 1, \\
         & D_{- 3 / 2}(C_3) = - 1, \\
         & D_{- 3 / 2}((C_3)^2) = 1,
        \end{align}
       \end{subequations}
       therefore,
       \begin{equation}
         D_{- 3 / 2} \equiv D_{+ 3 / 2}.
       \end{equation}
       In other words, the basis functions $\ket{+ \frac{3}{2}}$ and $\ket{- \frac{3}{2}}$ belong to the same IR.
       However, the presence of the TRS ${\cal T}$ gives rise to additional degeneracy, since the basis $\ket{- \frac{3}{2}}$ is linearly independent of $\ket{+ \frac{3}{2}}$.
 \item $s_z = + 1 / 2$ case. \\
       From Eq.~\eqref{eq:rotation_sz}, the 1D IR matrices $D_{+ 1 / 2}$ with the basis $\ket{+ \frac{1}{2}}$ are written by
       \begin{subequations}
        \begin{align}
         D_{+ 1 / 2}(E) &= 1, \\
         D_{+ 1 / 2}(C_3) &= e^{+ i\pi / 3}, \\
         D_{+ 1 / 2}((C_3)^2) &= e^{+ i 2\pi / 3}.
        \end{align}
       \end{subequations}
       Thus the Wigner criterion for the TRS ${\cal T}$ is calculated by
       \begin{align}
        W_{+ 1 / 2}^{\cal T} &= \frac{1}{3} \sum_{h \in H} z_{{\cal T} h, {\cal T} h} \chi[D_{+ 1 / 2}(({\cal T} h)^2)] \notag \\
        &= \frac{1}{3} ( - 1 - e^{i 2\pi / 3} + e^{i\pi / 3} ) = 0. \label{eq:Wigner_criterion_+1/2}
       \end{align}
       Equation~\eqref{eq:Wigner_criterion_+1/2} indicates that ${\cal T} \ket{+ \frac{1}{2}} \propto \ket{- \frac{1}{2}}$ has the nonequivalent rotational property to $\ket{+ \frac{1}{2}}$:
       \begin{subequations}
        \begin{align}
         & D_{- 1 / 2}(E) = 1, \\
         & D_{- 1 / 2}(C_3) = e^{- i\pi / 3}, \\
         & D_{- 1 / 2}((C_3)^2) = e^{- i2\pi / 3},
        \end{align}
       \end{subequations}
       therefore,
       \begin{equation}
         D_{- 1 / 2} \neq D_{+ 1 / 2}.
       \end{equation}
       Due to the bases of nonequivalent IRs $\ket{+ \frac{1}{2}}$ and $\ket{- \frac{1}{2}}$, the degeneracy is doubled by applying the TRS ${\cal T}$.
\end{enumerate}

\section{Orthogonality test}
\label{sec:orthogonality_test}
We explain the formulation of the orthogonality test for CS, used in Sec.~\ref{sec:gap_classification_topological}.
First, let $H$ a (unitary) crystal point group and $\alpha$ is a certain $d_\alpha$-dimensional IR of $H$ which has the basis functions $\psi_i$ ($i = 1, 2, \dots, d_\alpha$).
$\psi_i$ transforms under the symmetry operation $h \in H$ as
\begin{equation}
 h \psi_i = \sum_{j = 1}^{d_\alpha} \psi_j [D_\alpha(h)]_{ji},
\end{equation}
where $D_\alpha$ is a representation matrix of the IR $\alpha$.
Then, we consider the situation that the system has additional CS $\Gamma$: $H + \Gamma H$.
Let $\{z_{h_1, h_2}\} \in Z^2(H + \Gamma H, U(1))$ be a factor system of $H + \Gamma H$.
The orthogonality between $\{\psi_i\}$ and $\{\Gamma \psi_i\}$ is investigated in the following.

The basis $\Gamma \psi_i$ is transformed by $h \in H$ as
\begin{align}
 h (\Gamma \psi_i) &= \Gamma (\Gamma^{-1} h \Gamma) \psi_i \notag \\
 &= \sum_{j = 1}^{d_\alpha} (\Gamma \psi_j) z_{h, \Gamma} z_{\Gamma^{-1}, h \Gamma} [D_\alpha(\Gamma^{-1} h \Gamma)]_{ji}, \notag \\
 &= \sum_{j = 1}^{d_\alpha} (\Gamma \psi_j) \frac{z_{h, \Gamma}}{z_{\Gamma, \Gamma^{-1} h \Gamma}} [D_\alpha(\Gamma^{-1} h \Gamma)]_{ji}, \label{eq:IR_matrix_CS}
\end{align}
where we use the 2-cocycle condition
\begin{equation}
 z_{h_2, h_3} z_{h_1 h_2, h_3}^{-1} z_{h_1, h_2 h_3} z_{h_1, h_2}^{-1} = 1,
\end{equation}
for $h_1, h_2, h_3 \in H + \Gamma H$.
We remark that $H$ is not changed under the CS: $\Gamma^{-1} H \Gamma = H$.
In other words, the representation matrix of $H$ with the bases $\{\Gamma \psi_i\}$ is given by $\frac{z_{h, \Gamma}}{z_{\Gamma, \Gamma^{-1} h \Gamma}} D_\alpha(\Gamma^{-1} h \Gamma)$.
Next, we recall the orthogonality relation between two IR matrices $D_\alpha$ and $D_\beta$~\cite{Inui-Tanabe-Onodera, Shiozaki2018_arXiv},
\begin{equation}
 \frac{1}{|H|} \sum_{h \in H} [D_\alpha(h)]_{ij} [D_\beta(h)^*]_{kl} = \frac{1}{d_\alpha} \delta_{\alpha\beta} \delta_{ik} \delta_{jl}. \label{eq:orthogonality_IR_matrix}
\end{equation}
By taking $i = j$ and $k = l$, we calculate the summation of Eq.~\eqref{eq:orthogonality_IR_matrix} over $i$ and $k$:
\begin{align}
 \frac{1}{|H|} \sum_{h \in H} \sum_{i = 1}^{d_\alpha} [D_\alpha(h)]_{ii} \sum_{k = 1}^{d_\beta} [D_\beta(h)^*]_{kk} &= \frac{\delta_{\alpha\beta}}{d_\alpha} \sum_{i = 1}^{d_\alpha} \sum_{k = 1}^{d_\alpha} \delta_{ik}, \notag \\
 \therefore \frac{1}{|H|} \sum_{h \in H} \chi[D_\alpha(h)] \chi[D_\beta(h)^*] &= \delta_{\alpha\beta}. \label{eq:orthogonality_IR_char}
\end{align}
Finally, Eqs.~\eqref{eq:IR_matrix_CS} and \eqref{eq:orthogonality_IR_char} lead to the orthogonality test between $\{\psi_i\}$ and $\{\Gamma \psi_i\}$,
\begin{align}
 & \frac{1}{|H|} \sum_{h \in H} \frac{z_{h, \Gamma}^*}{z_{\Gamma, \Gamma^{-1} h \Gamma}^*} \chi[D_\alpha(h)] \chi[D_\alpha(\Gamma^{-1} h \Gamma)^*] \notag \\
 & \quad = \begin{cases}
            1 & (\text{$\{\psi_i\}$ and $\{\Gamma \psi_i\}$ are equivalent}), \\
            0 & (\text{$\{\psi_i\}$ and $\{\Gamma \psi_i\}$ are nonequivalent}),
           \end{cases}
\end{align}
which is nothing but Eq.~\eqref{eq:orthogonality_test_G}.


%

\end{document}